\documentclass{llncs}
\usepackage{graphicx}
\usepackage{caption}
\usepackage{subfig}
\usepackage{amsmath}
\usepackage[utf8]{inputenc}
\usepackage[english]{babel}
\usepackage{algorithm}
\usepackage{amsfonts}
\usepackage[noend]{algpseudocode}
\usepackage{geometry}
\begin{document}
\title{A Theoretical Model for Understanding the Dynamics of Online Social Networks Decay}
\author{Mohammed Abufouda}
\institute{Computer Science Department, University of Kaiserslautern,\\Gottlieb-Daimler-Str. 48, 67663, Kaiserslautern, Germany\\abufouda@cs.uni-kl.de}
\maketitle

\begin{abstract}
Online social networks represent a main source of communication and information exchange in today's life. They facilitate exquisitely news sharing, knowledge elicitation, and forming groups of same interests. Researchers in the last two decades studied the growth dynamics of the online social networks extensively questing a clear understanding of the behavior of humans in online social networks that helps in many directions, like engineering better recommendation systems and attracting new members. However, not all of social networks achieved the desired growth, for example, online social networks like MySpace, Orkut, and Friendster are out of service today. In this work, we present a probabilistic theoretical model that captures the dynamics of the social decay due to the inactivity of the members of a social network. The model is proved to have some interesting mathematical properties, namely \textit{submodularity}, which imply achieving the model optimization in a reasonable performance. That means the maximization problem can be approximated within a factor of $(1-1/e)$ and the minimization problem can be achieved in polynomial time.
\end{abstract}

\section{Introduction}
\label{sec:Introduction}
Today's online social networks represent a main source of communication and information exchange among people all over the world. Many online social networks have proven their superiority, like Facebook, Twitter, and Linkedin, in connecting people and facilitating an exquisite new medium for sharing news, forming groups of people of the same interests, and eliciting knowledge. The growth of these networks\footnote{ For example, the growth of  Facebook: http://www.statista.com/statistics/264810/number-of-monthly-active-facebook-users-worldwide/} in terms of user activity shows that these online social networks have become a vital part in today's human activities. Thus, studying the human behavior in these networks has become a necessity in order to understand the behavior of humans and their interactions in these networks.\\

\textbf{Social network growth:} One well studied aspect of online social networks dynamics is the \textit{growth} phenomenon of a network. The work by {Barab{\'a}si et al.~\cite{Barabasi1999} presented a simple model for understanding the growth dynamics of a network, namely \textit{Preferential Attachment} which is a rich get richer model. Jin et al.~\cite{jin2001} noticed that the model by {Barab{\'a}si et al.~\cite{Barabasi1999} and other similar models, like the work by Dorogovtsev et al.~\cite{dorogovtsev2000} for modeling the growth of random networks, are not suitable to understand the growth dynamics of social networks. Thus, they provided a model that considers the specialty of social networks like not having, seemingly, a power law distribution and the existence of clustering property~\cite{jin2001}. With the avaliability of the online datasets, Newman~\cite{newman2001} studied empirically the growth of social networks using the scientific collaboration networks against the \textit{Preferential Attachment}~\cite{Barabasi1999} model. Then, Leskovec et al.~\cite{leskovec2005} came with law-like properties of networks over time, namely, densification of networks and shrinking diameter of networks over time, assuming that the dynamics is growth dynamics.\\
The previous work and the avaliability of rich datasets pushed the research to an in-depth investigation of the properties of the networks over time. Kumar et al.~\cite{Kumar2006} studied the growth of a large social network in terms of network component analysis, Kossinets et al.~\cite{kossinets2006} studied the tie formation process within the social networks that is affected by internal and external factors, and Capocci et al.~\cite{capocci2006} studied the statistical properties of and the growth characteristics of Wikipedia collaboration social networks. Likewise, Backstrom et al.~\cite{backstrom2006} studied empirically how groups are formed and evolved over time in MySpace social networks and Mislove et al.~\cite{mislove2008} provided a study for the growth of Flicker social network.\\

\textbf{Social network decay:} Even though there are many successful social networks, the evolution of social network in general is not restricted to a growth process. In the last decade, it has been noticed that some of the online social networks were closed after a huge abandon or inactivity of their members. Online social networks, like Friendsfeed, Friendster, MySpace, Orkut, and many websites of the Stack Exchange platform, are now out of service, where some of these websites, Orkut and Myspace, were a rich environment to study the social growth dynamics~\cite{ahn2007} just a decade ago. The decay of these networks poses many questions about the reasons behind their fall down. Garcia et al.~\cite{Garcia2013} and Chhabra et al.~\cite{Chhabra2014} studied the static properties of Friendster and MySpace, respectively, in order to understand the network-related properties of these networks as an example of a decayed network. Recent studies by Malliaros et al.~\cite{malliaros2013} and Bhawalkar et al.~\cite{bhawalkar2015} provided theoretical models for understanding the social engagement in online social networks with a potential to predict social inactivity.\\
While investigating the reasons behind the inactivity of the members of the online social networks is not in the scope of this work, some recent studies proposed some answers~\cite{stieger2013,kordestani2015}, suggesting that the main reason behind this decay is the inactivity of the members of the online social networks.\\
Building a sound understanding of the decay dynamics of networks requires not only studying the static properties of these networks, but also requires investigating their behavior over time, and this is what we are interested in.
Here, we consider the Stack Exchange websites that were closed after a period of time due to the lack of enough activity required to keep the website alive\footnote{A list of all closed stack exchanges websites can be found here: http://bit.ly/2bVeukz}. The closed websites are an example of the social network decay, where we model the members of a closed website as the nodes of the network and an edge exists between any two nodes if they post, comment, or answer in the same question in the websites.

\textbf{The decayed websites of Stack Exchange:} Before proceeding, we will present some analysis and comparisons between the decayed websites (websites that are closed) and alive ones. 
Figure~\ref{fig:comments} shows the distribution of the number of users comments for alive and decayed websites. The figure shows that the decayed websites clearly have different distribution characteristics with a low mean and low standard deviation. A similar behavior is found in Figure~\ref{fig:reputation} and Figure~\ref{fig:upvotes} that represents the distribution of users total received \textit{Reputation} and \textit{Upvotes}, respectively. These two properties reflect the level of knowledge and experience that the members of a website have. For the decayed websites, it is clear that, on average, the members have much less reputation and upvotes than those in the alive websites. This may indicate that one reason behind closing websites is the lack of experienced members.\\
\begin{figure}[htbp]
\centering
\includegraphics[width=10cm]{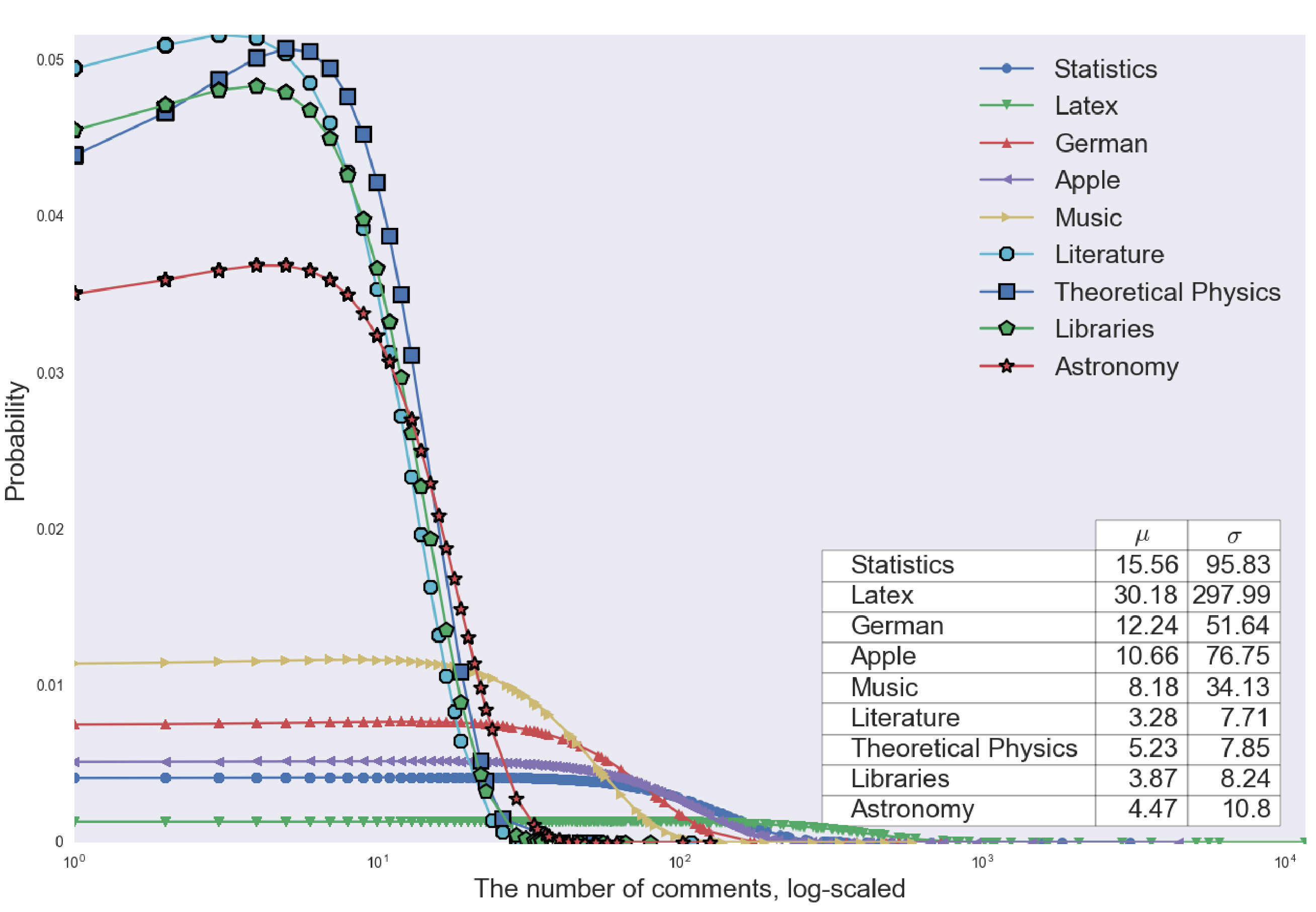}
\caption{(Color Online) The probability distribution of the comments for different websites. Markers with bold boarders are decayed websites, $\mu$ is the mean, and $\sigma$ is the standard deviation. From the figure it is clear that the decayed networks have different distribution properties from the other alive networks.} 
\label{fig:comments}
\end{figure} 
All of the three figures, Figures~\ref{fig:comments},~\ref{fig:reputation}, and~\ref{fig:upvotes} show that there is less social activity in the decayed websites, which may be used as an indication for studying the future of the alive websites. However, understanding the decay dynamics of the decayed websites requires a deeper investigation and modeling for the nature of the interaction among the members.

Our approach to better understand what happen during the decay process is to make a network representation of the members interactions, like comments, upvotes, and posts, as networks. Then, we build a network based model for modeling the decay process. Figure~\ref{fig:networksDecay} shows the real-data network representation of the \textit{Startup Business} decayed website over time\footnote{The material related to this paper and more visualization videos for many decayed websites are available here: http://www.abufouda.com/research/decay-dynamics-in-online-social-networks/}. In the figure, there is clear decrease in the number of nodes and the number of edges in the network over time, which is another indication about the decay of the social interaction over time in the decayed websites.

\begin{figure}[htbp]
\centering
\includegraphics[width=12cm]{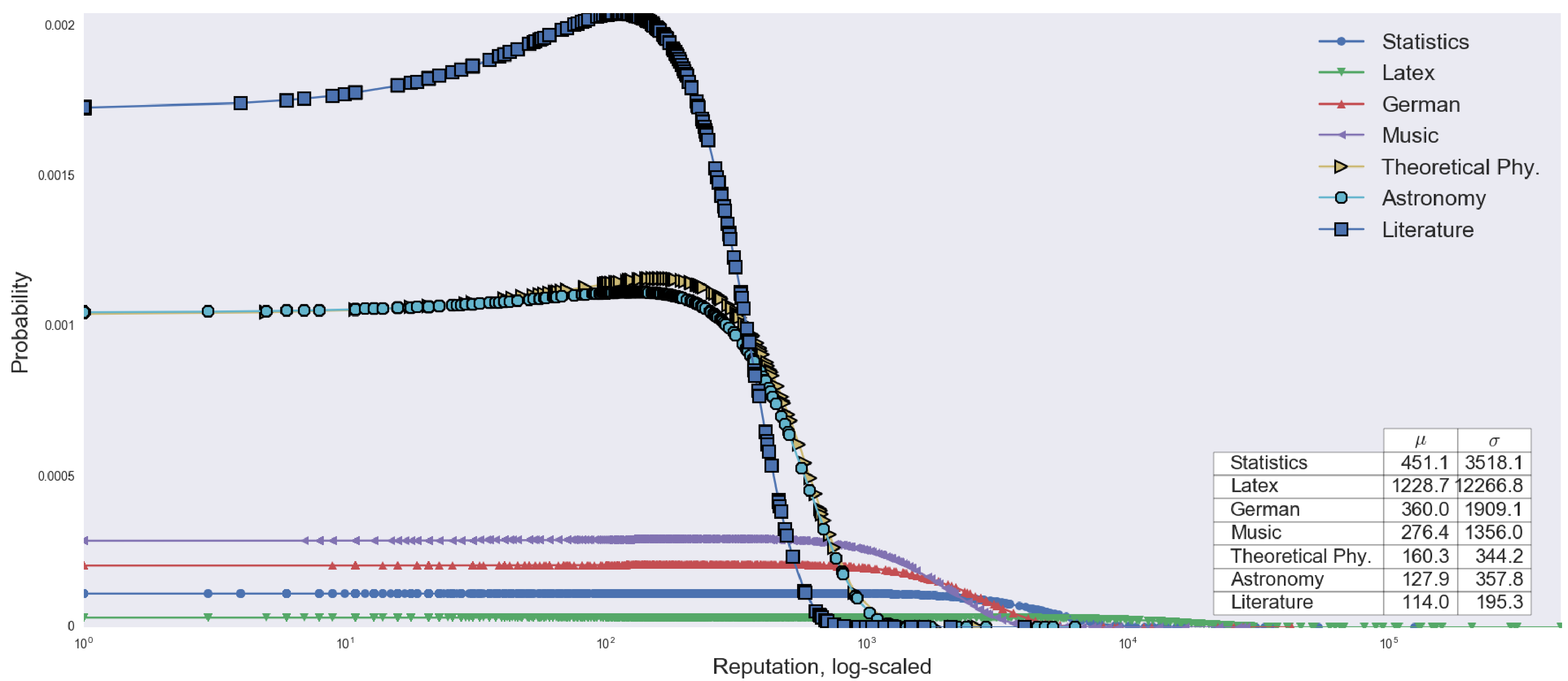}
\caption{(Color Online) The probability distribution of users "Reputation" for different websites. Markers with bold boarders are decayed websites, $\mu$ is the mean, and $\sigma$ is the standard deviation.}
\label{fig:reputation}
\includegraphics[width=12cm]{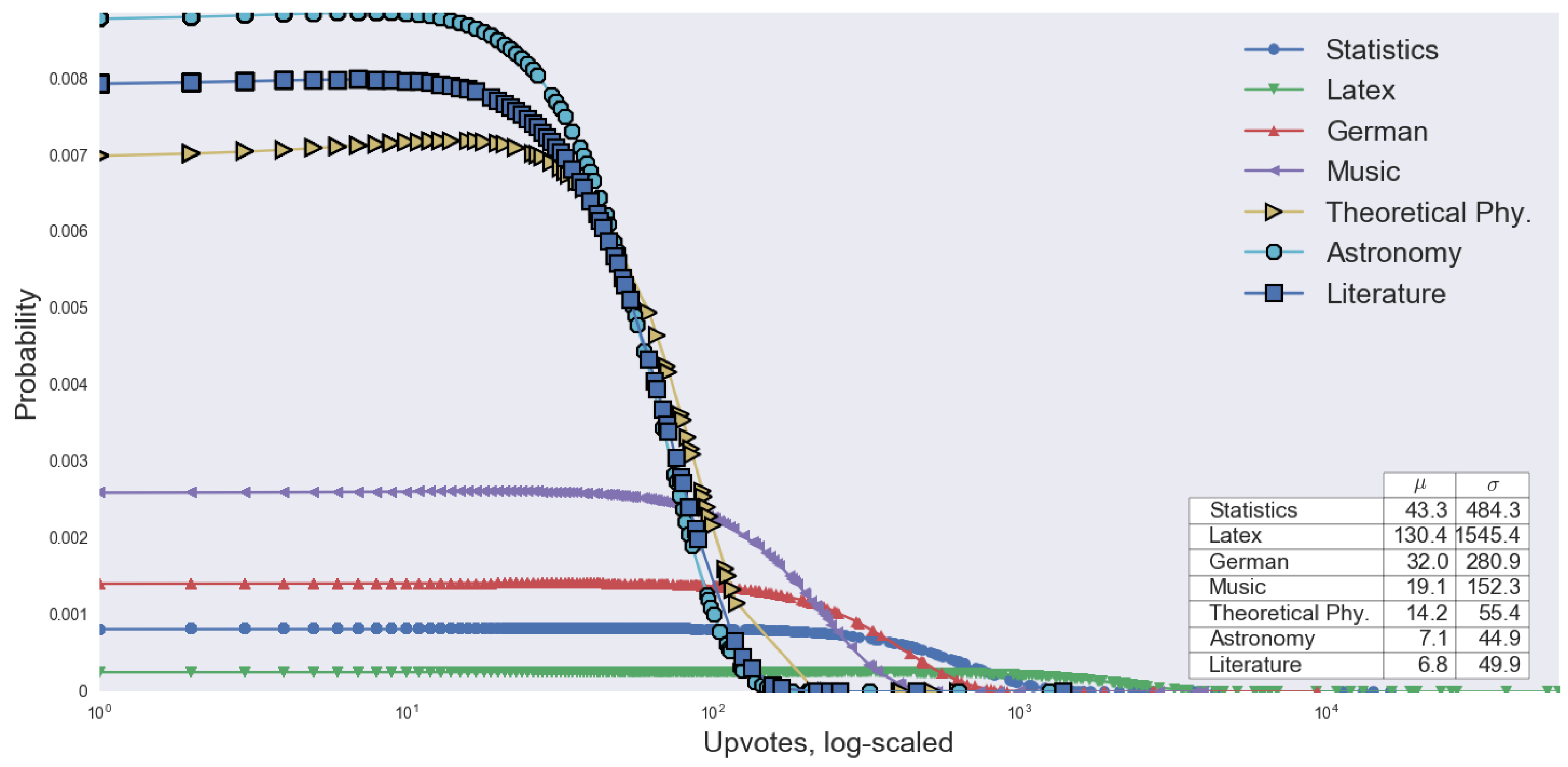}
\caption{(Color Online) The probability distribution of users "Upvotes" for different websites. Markers with bold boarders are decayed websites, $\mu$ is the mean, and $\sigma$ is the standard deviation.}
\label{fig:upvotes}
\end{figure}

\begin{figure}
\centering
\parbox{4.51cm}{
\includegraphics[width=4.51cm]{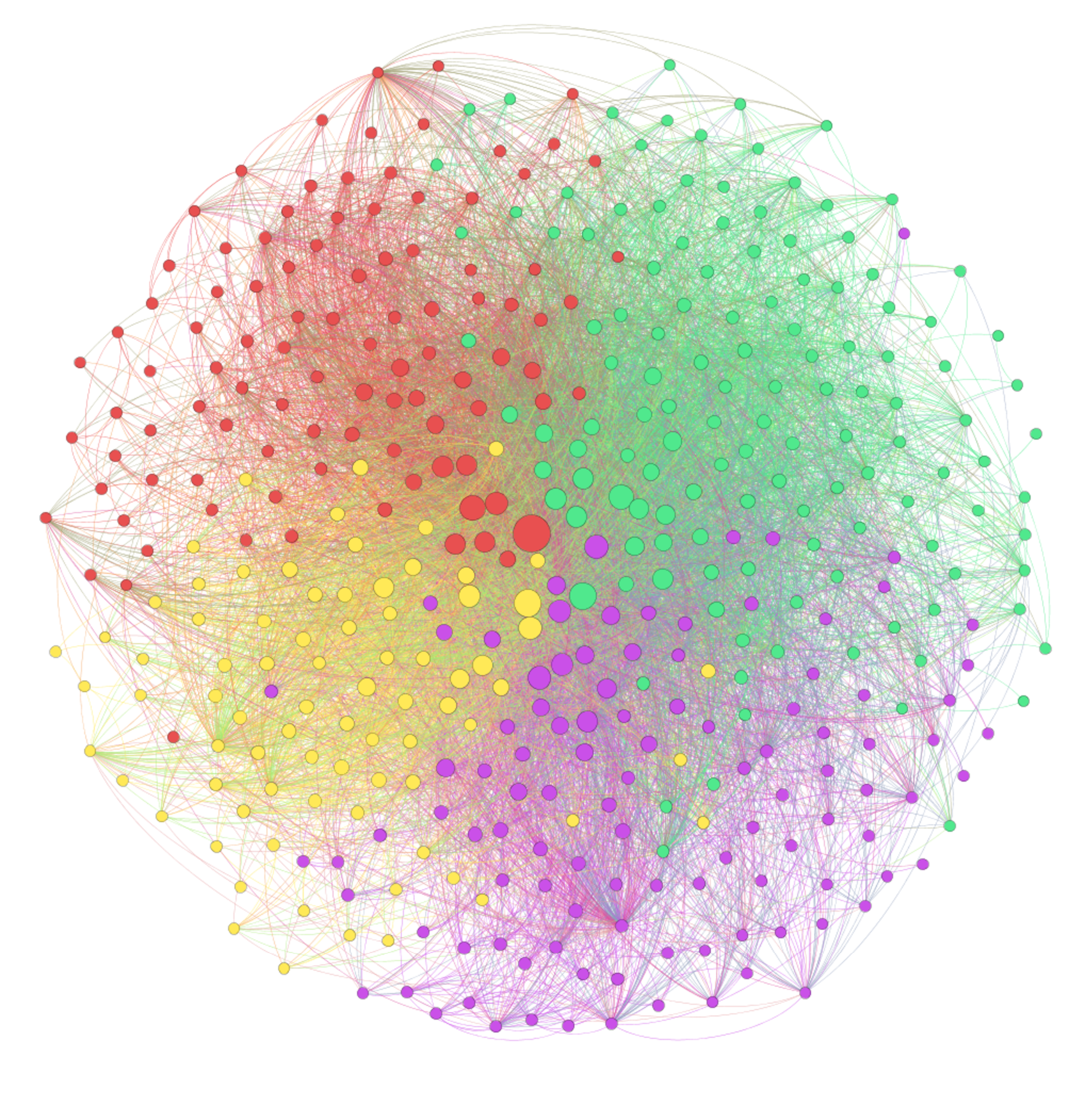}
\caption*{Oct-2009}
\label{fig:toy1}}
\qquad
\begin{minipage}{4.51cm}
\includegraphics[width=4.51cm]{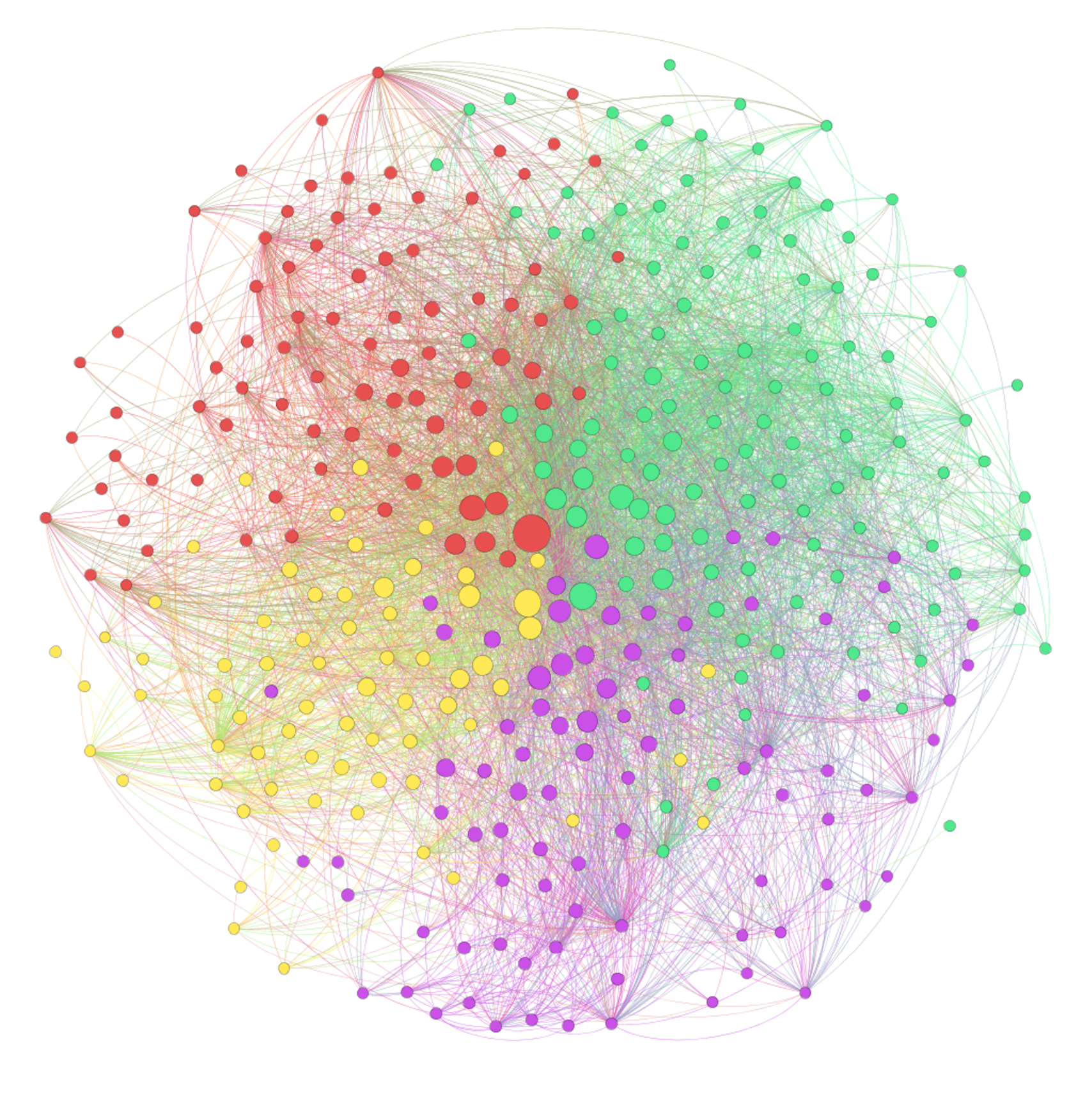}
\caption*{Dec-2009}
\label{fig:toy2}
\end{minipage}
\qquad
\begin{minipage}{4.51cm}
\includegraphics[width=4.51cm]{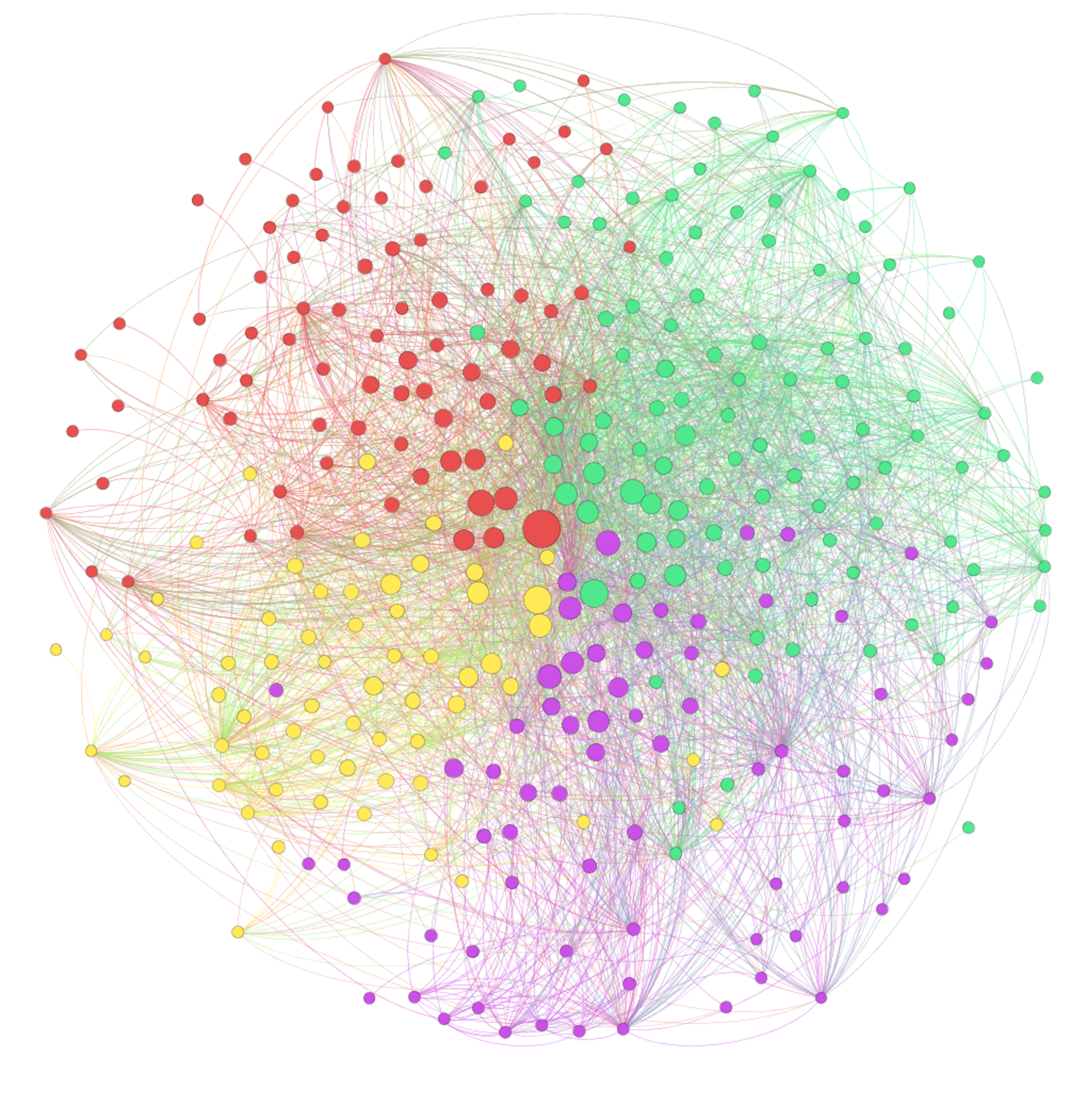}
\caption*{Feb-2010}
\label{fig:toy2}
\end{minipage}
\qquad
\begin{minipage}{4.51cm}
\includegraphics[width=4.51cm]{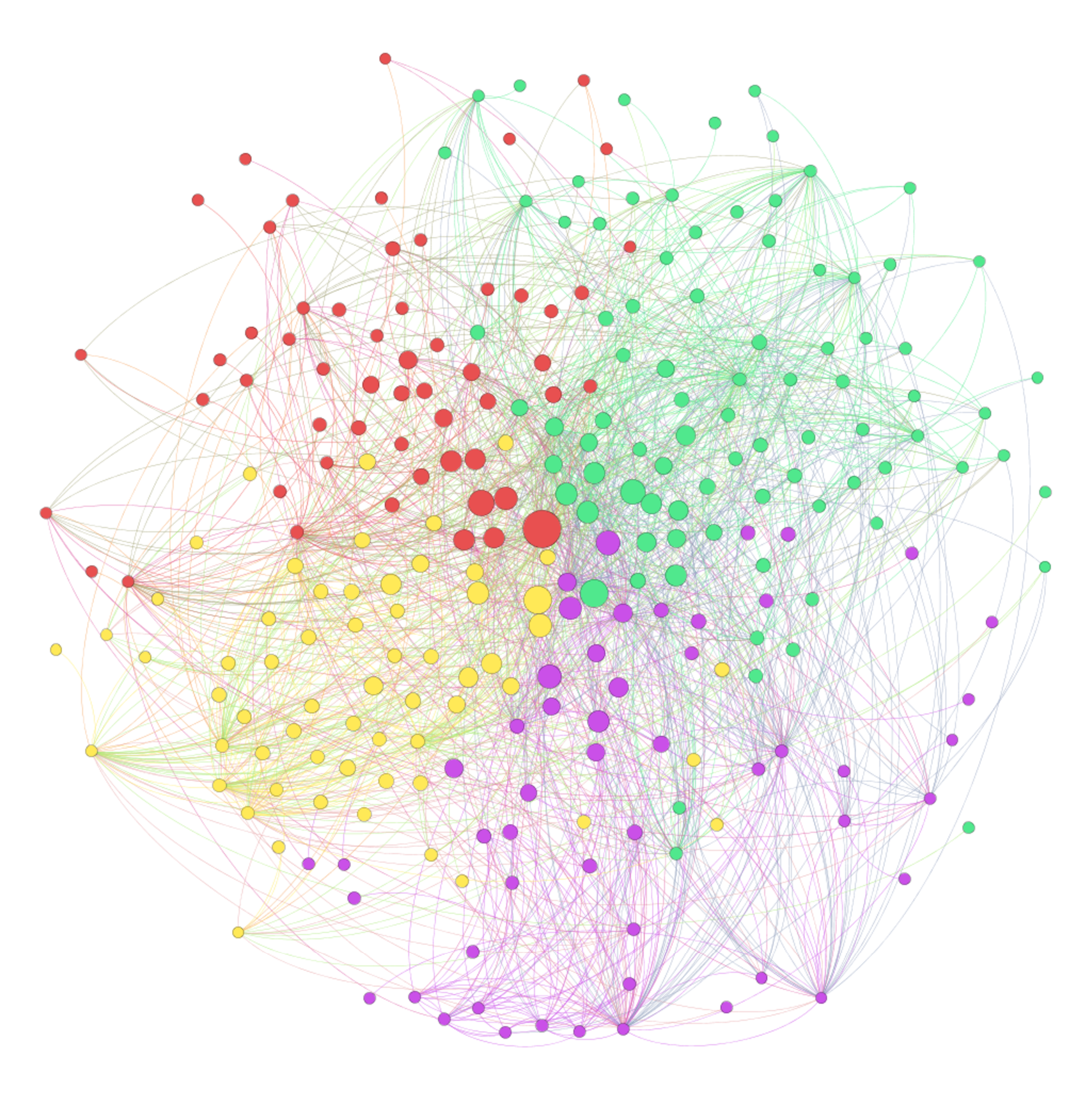}
\caption*{Apr-2010}
\label{fig:toy2}
\end{minipage}
\qquad
\begin{minipage}{4.51cm}
\includegraphics[width=4.51cm]{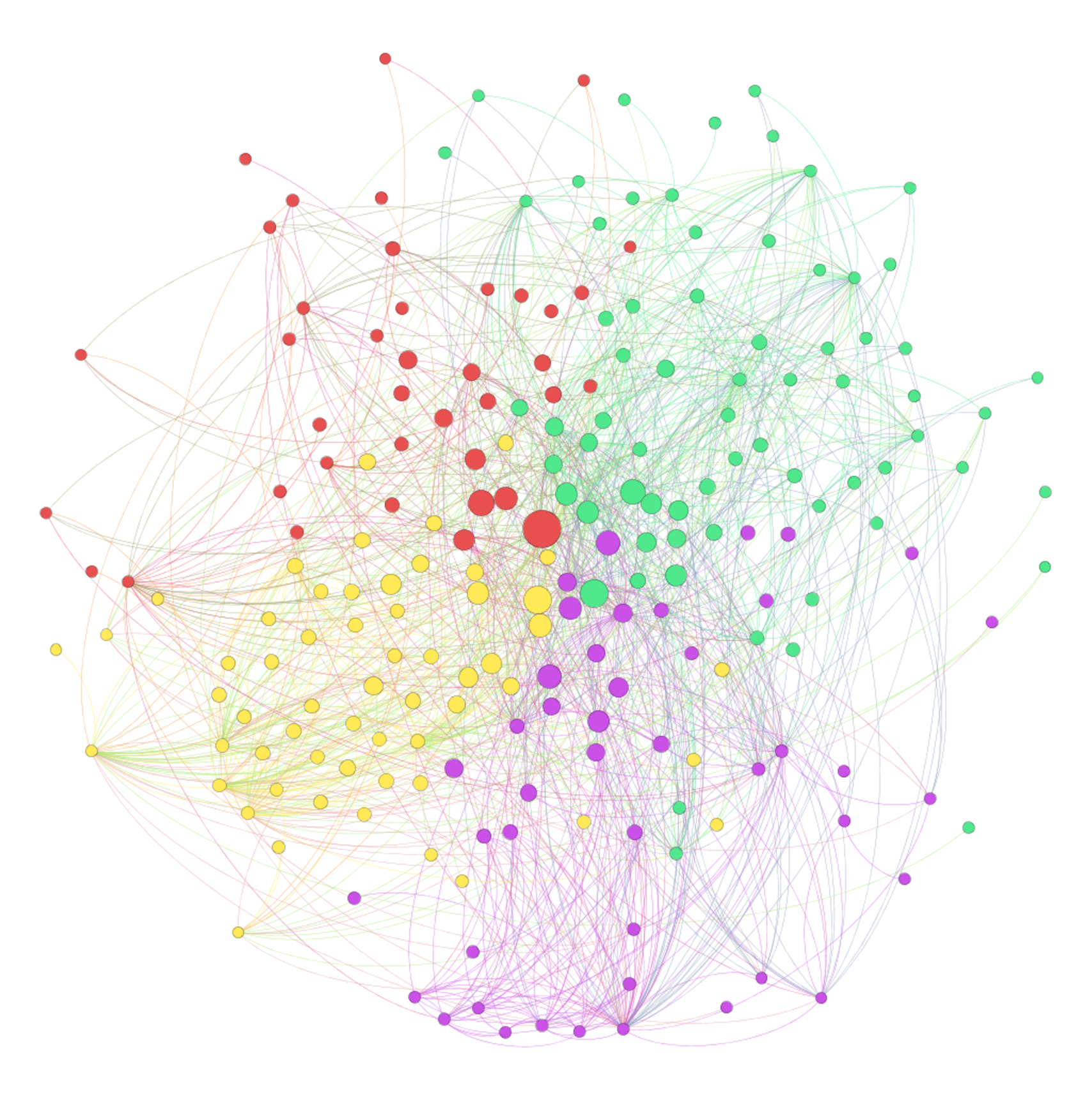}
\caption*{Jun-2010}
\label{fig:toy2}
\end{minipage}
\qquad
\begin{minipage}{4.51cm}
\includegraphics[width=4.51cm]{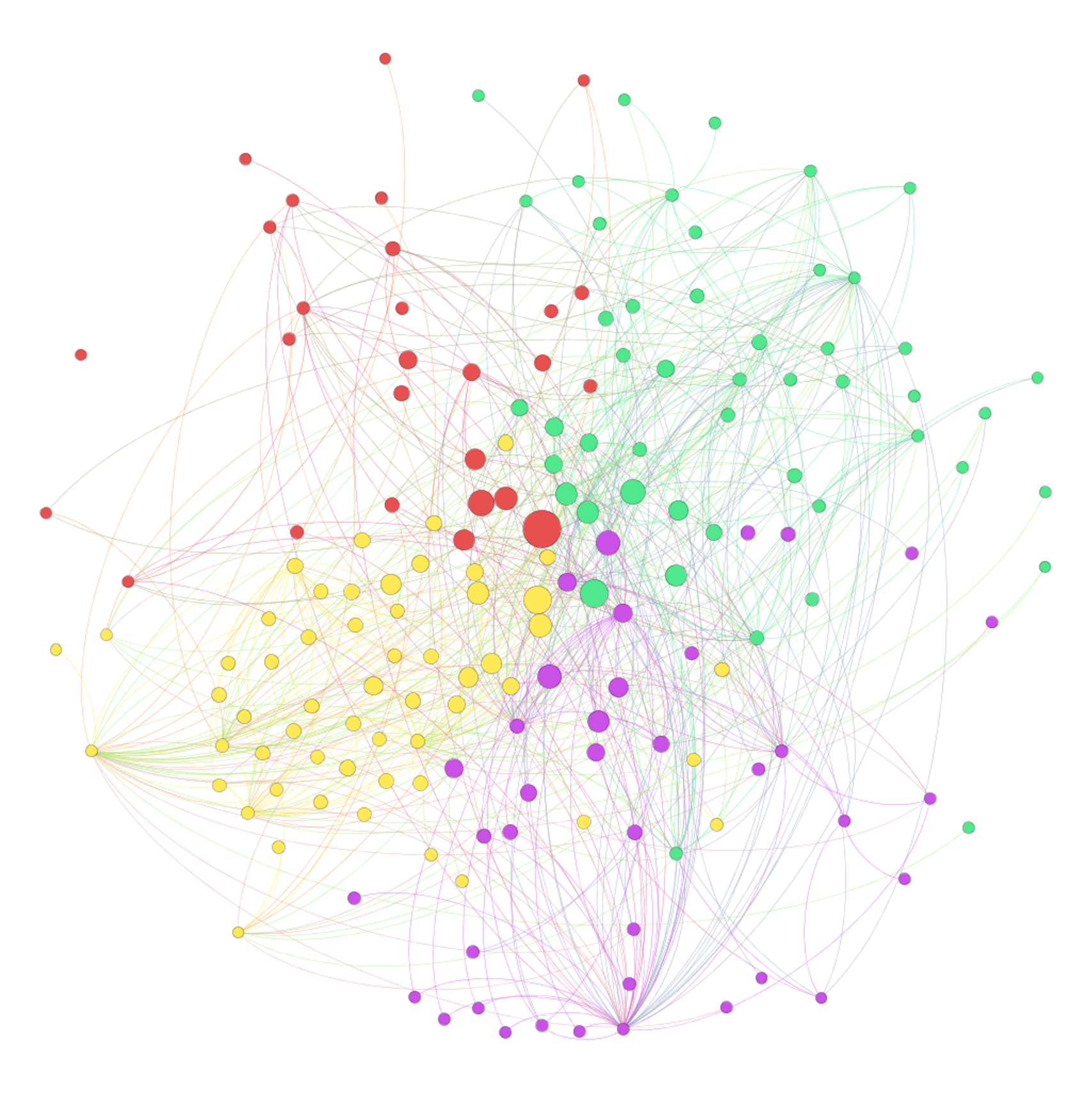}
\caption*{Aug-2010}
\label{fig:toy2}
\end{minipage}
\qquad
\begin{minipage}{4.51cm}
\includegraphics[width=4.51cm]{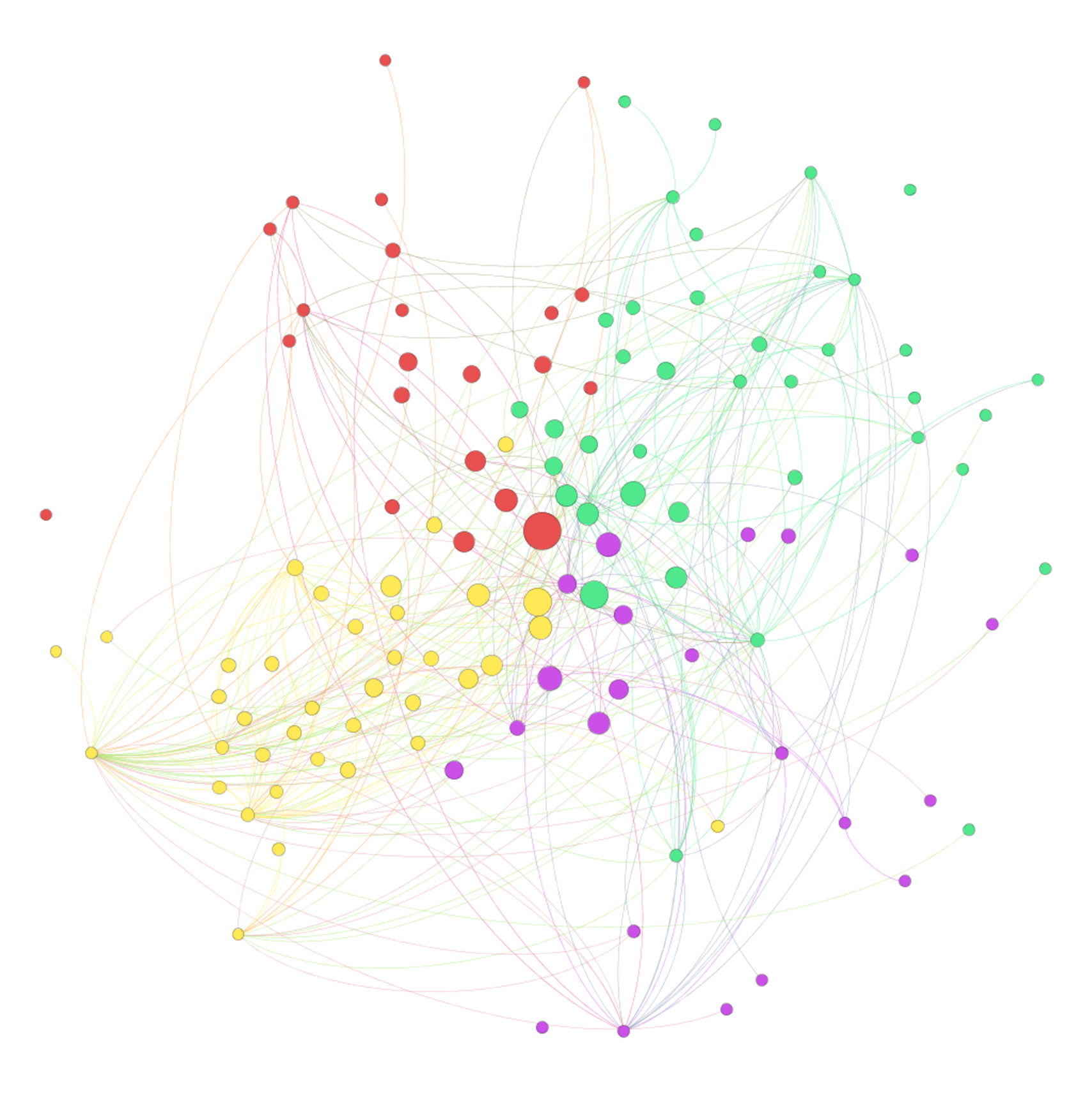}
\caption*{Oct-2010}
\label{fig:toy2}
\end{minipage}
\qquad
\begin{minipage}{4.51cm}
\includegraphics[width=4.51cm]{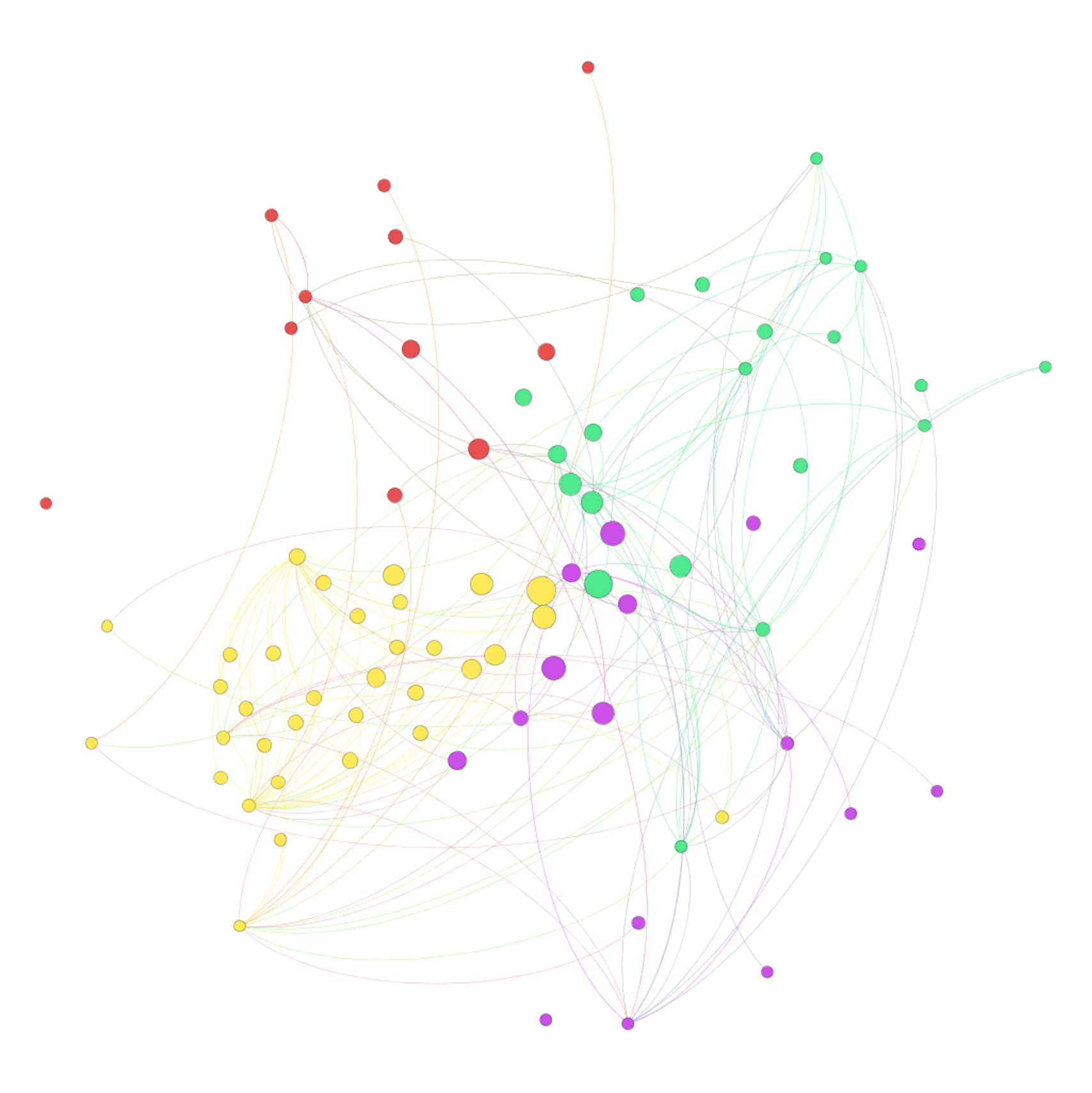}
\caption*{Jan-2011}
\label{fig:toy2}
\end{minipage}
\qquad
\begin{minipage}{4.51cm}
\includegraphics[width=4.51cm]{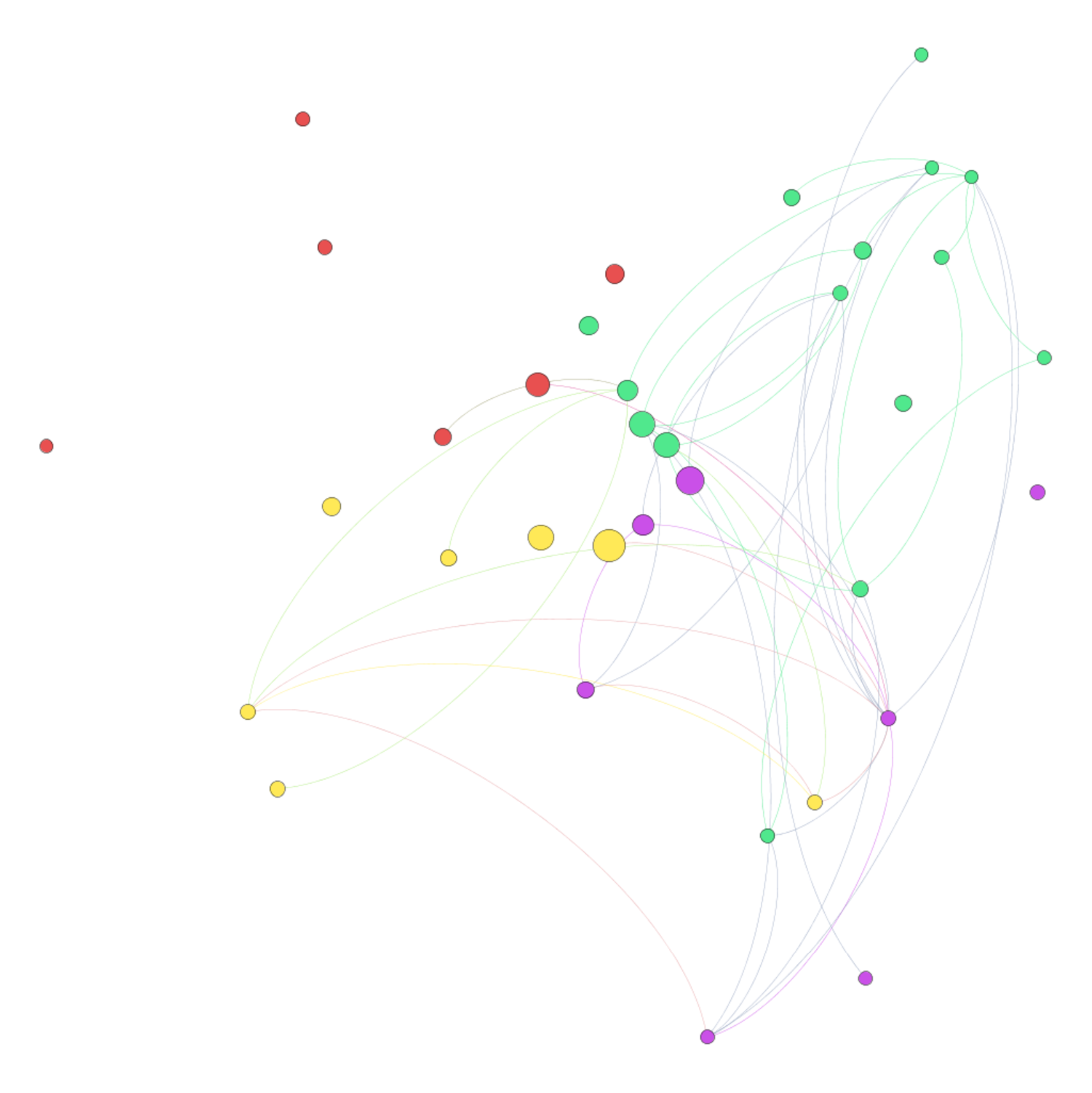}
\caption*{May-2011}
\label{fig:toy2}
\end{minipage}
\qquad
\begin{minipage}{4.51cm}
\includegraphics[width=4.51cm]{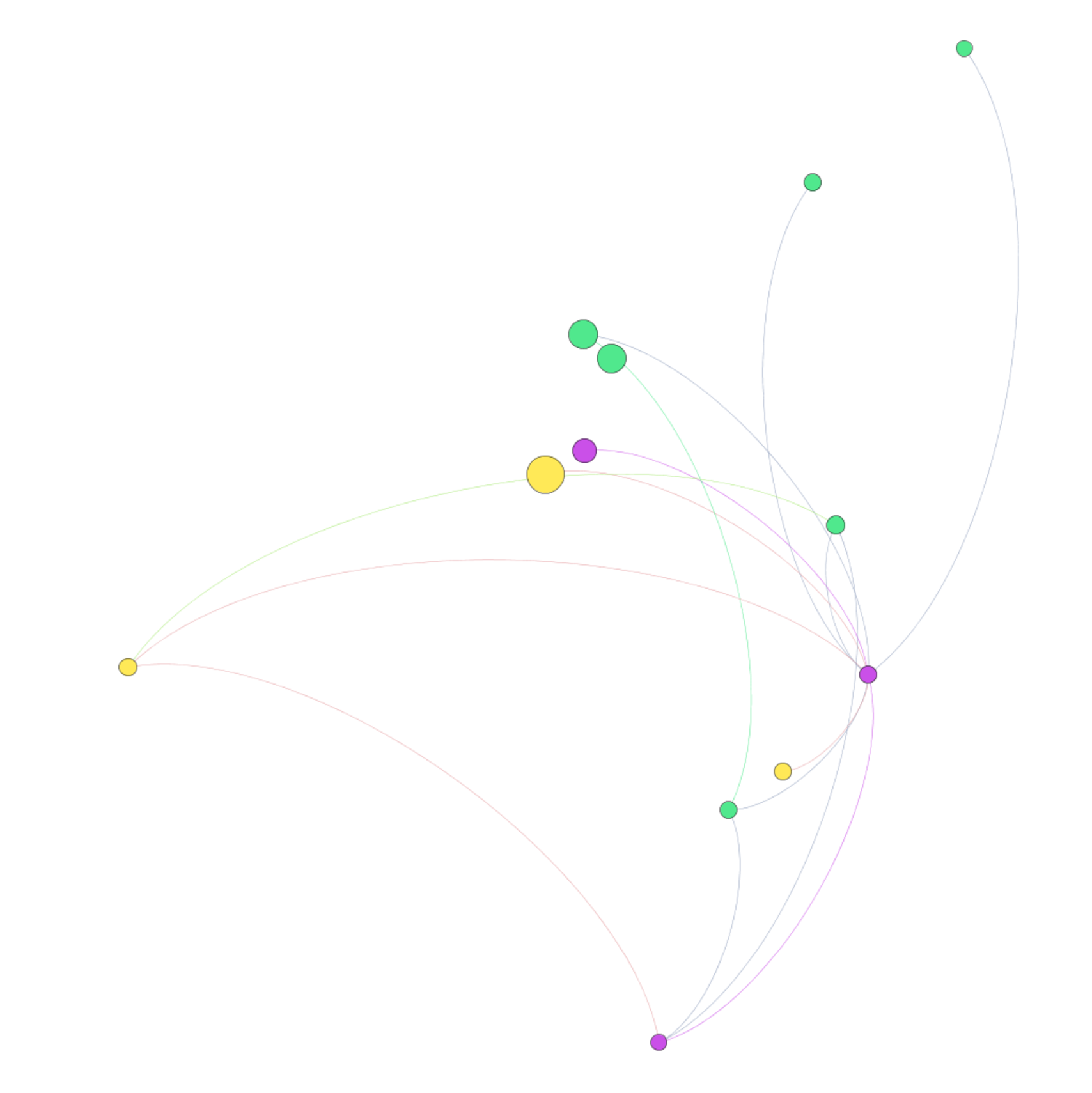}
\caption*{Nov-2011}
\label{fig:toy2}
\end{minipage}
\label{fig:toy2}
\caption{(Color Online) The network representation of the real data of the "Business Startups" website social network between Oct-2009 and Nov-2011. The networks (nodes are the members and edges represent interaction between any two members) were colored by clusters and having a node size directly proportional to its degree. For the sake of simplicity and for a better visualization, we restricted the networks to core members who registered in during the first four months of the website. The $10$ networks in the figure show a clear decay due to the inactivity of the members, leading to the close of the website. The figure also shows that a network representation of the interaction is a good abstraction (model) choice to capture the decay of the interaction of among the members of the website.}
\label{fig:networksDecay}
\end{figure}
\vspace*{-0.5cm}
\textbf{Our contribution:} Modeling a human behavior (the online social decay in our case) is not an easy task. Thus, we are interested in modeling the consequences of this behavior. Having that said, we are answering \textit{how}, not \textit{why}, the social decay happen. We are not interested in knowing the reasons behind social decay, instead we are interested in modeling the consequences (the effects) of this behavior, which may help in understanding the reasons of social decay (the causes).\\
In this work, we provide a probabilistic model for understanding the social decay phenomenon in online social networks. We think that providing a mechanistic model will be a useful tool to better understand what we call the \textit{decay dynamics}, which will certainly enhance our understanding of the human behavior in online social networks. Although the human behavior has some regularities, we strongly believe that it is not deterministic, unlike the physical systems. Thus, a probabilistic modeling is a good way to mitigate the non-determinism of human behavior. Our contribution in this work is split into:
\begin{itemize}
\item A probabilistic model for capturing the social decay. The model represents a \textit{step by step} mechanistic model that captures the leave of a node in a network under decay and quantify the consequent properties of the network after its leave.
\item Measures for understanding the network resilience to node leaves and the influence of a node leave on both the neighbors and the network.
\item Theoretical proof of the submodularity of the model that leads to viable optimization. Being submodular, the maximization problem of the model can be approximated to a reasonable factor, and the minimization problem can be done in polynomial time.
\end{itemize}

\section{Model and notations}
\label{sec:modelandnotations}
A network $G=(V,E)$ is a tuple of two sets $V$ and $E$, where $V$ is the set of nodes and $E$ is the set of edges such that an undirected edge $e$ is defined as $e=\{u,v\} \in E$, where $u,v \in V$. As we consider the a dynamic system, the notation $G^t$ is a network at time $t$.
We assume that every node $w \in V$ has an initial \textit{Leave Probability} $\pi_w^0$ which determines whether the node $w$ will leave\footnote{The word "leave" means any kind of social inactivity like deleting an online account, deactivating it, or being inactive.} at time $1$, and generally at $t+1$, or not. If a node $w$ did not leave at $t+1$, i.e., $w \in V(G^{t+1})$, then its current leave probability, $\pi_w^t$, will be increased depending on its neighbors who left at $t-1$. The details of this process are described in the following sections.
The notion $l(w)$ denotes the time at which the node $w$ left the network.
\begin{definition}
A dynamic network $G$ is called a "Decaying Network" if $|E(G)^{t-1}| \geq |E(G)^{t}|$, $|V(G)^{t-1}| \geq |V(G)^{t}|$, and $V(G)^t \subseteq V(G)^{t-1}, \forall t > 0$.
\end{definition}
We assume the model starts with a \textit{Decaying Network}, where neither new nodes nor new edges appear in the network, and thus the network dynamics are restricted to node and link removal. Even though this assumption sounds unrealistic for many networks, but we emphasis that it is only unrealistic for networks with growth dynamics, not for the networks under decay which is what we are modeling here.
The main idea of the model is that the leave of a node will increase the leave probability of its adjacent nodes (Figure~\ref{fig:toy1} gives an example).

\begin{figure}[htbp]
\centering
\parbox{4.8cm}{
\includegraphics[width=4.8cm]{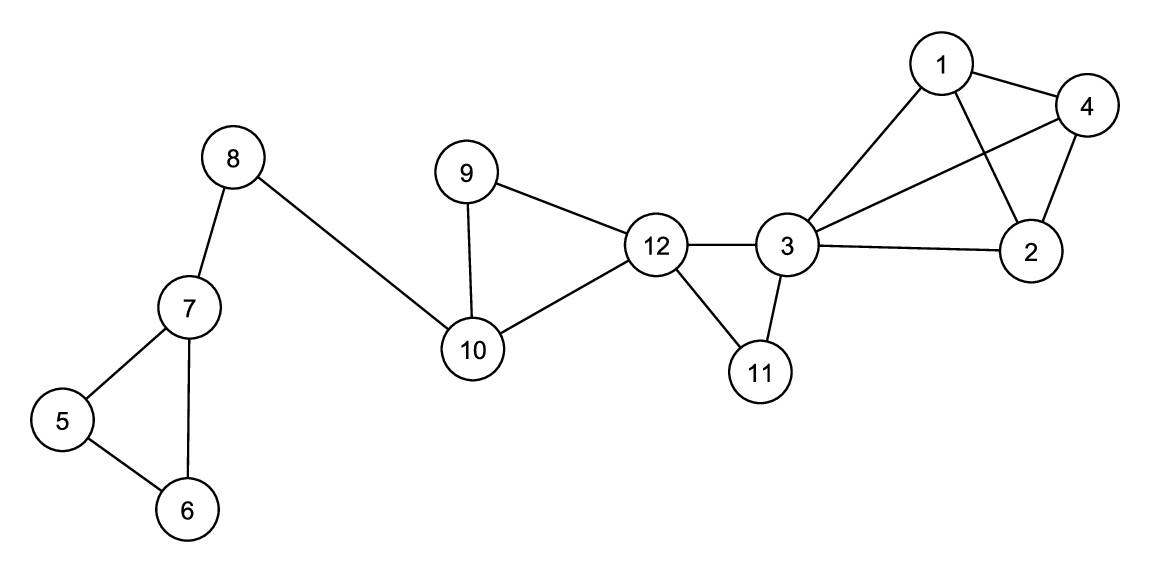}
\caption*{$t=0$}}
\qquad
\begin{minipage}{4.8cm}
\includegraphics[width=4.8cm]{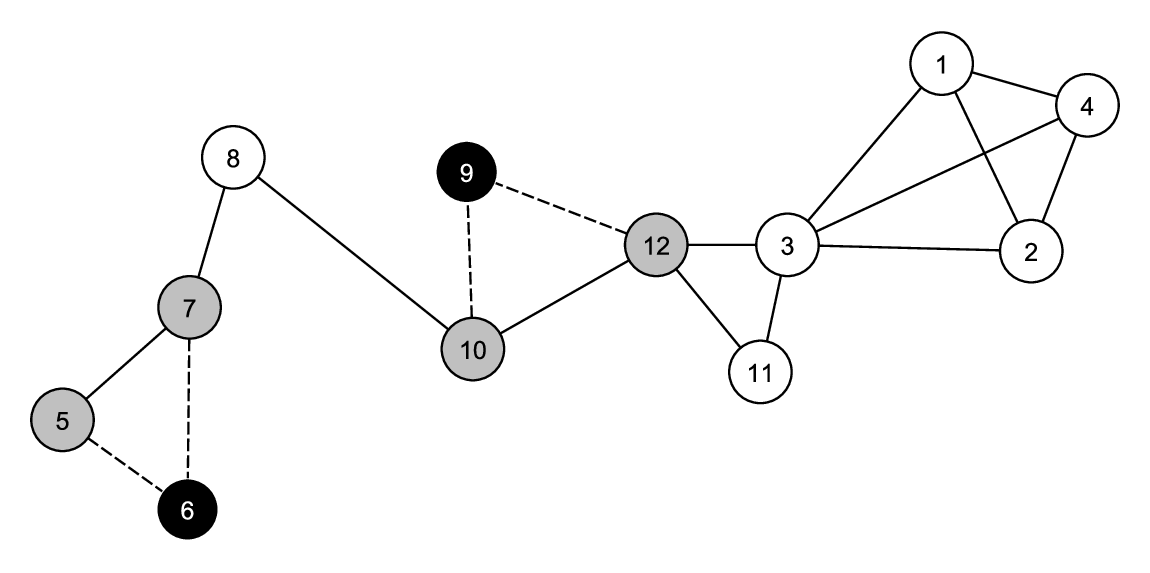}
\caption*{$t=1$}
\end{minipage}
\qquad
\begin{minipage}{4.8cm}
\includegraphics[width=4.8cm]{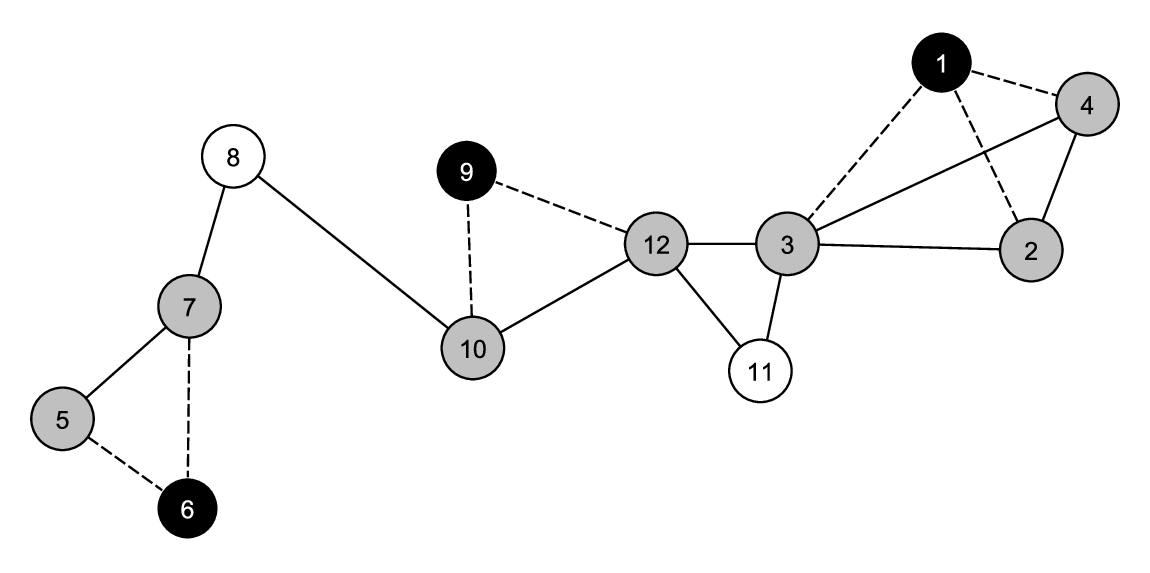}
\caption*{$t=2$}
\end{minipage}
\qquad
\begin{minipage}{4.8cm}
\includegraphics[width=4.8cm]{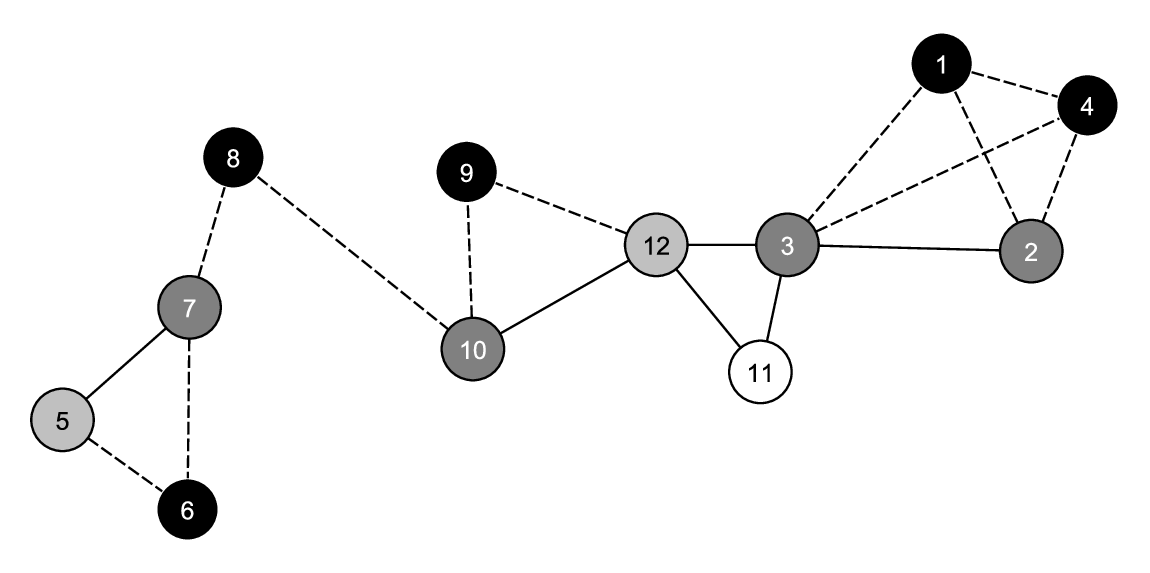}
\caption*{$t=3$}
\end{minipage}
\qquad
\begin{minipage}{4.8cm}
\includegraphics[width=4.8cm]{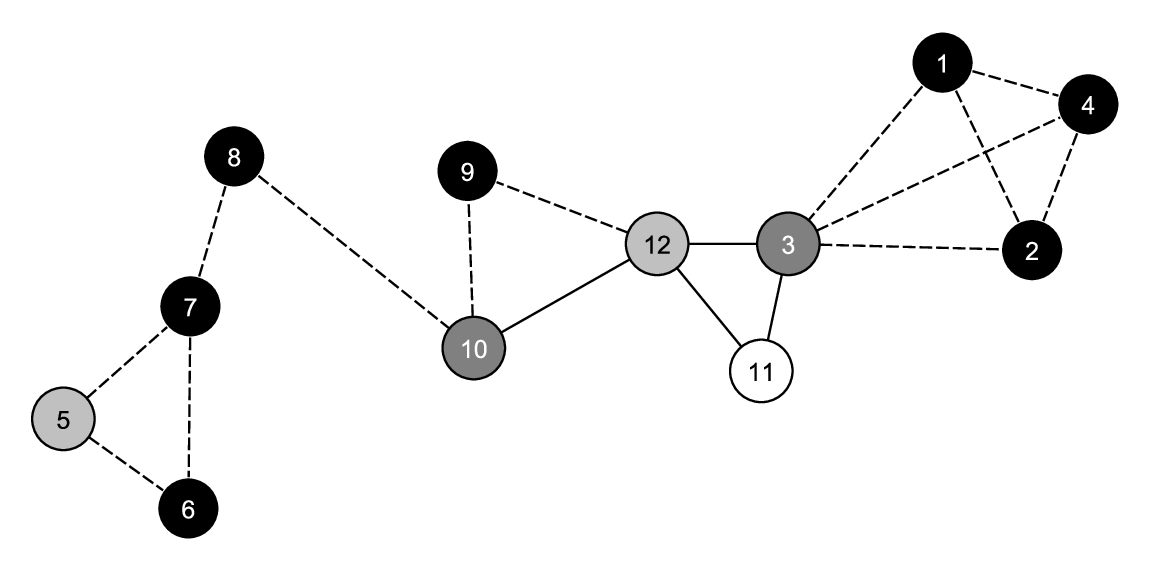}
\caption*{$t=4$}
\end{minipage}
\qquad
\begin{minipage}{4.8cm}
\includegraphics[width=4.8cm]{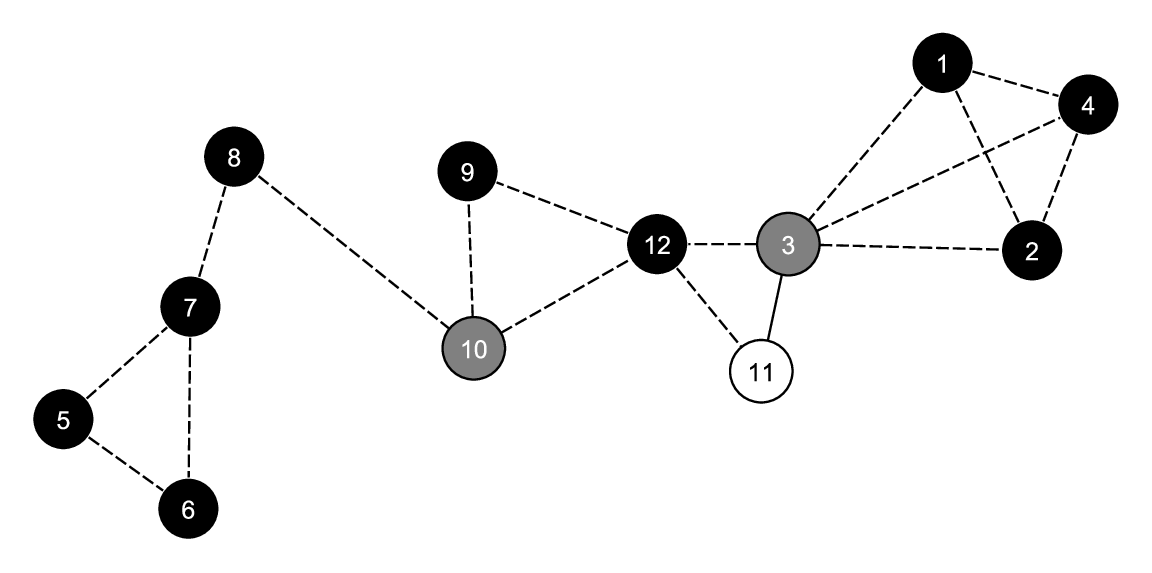}
\caption*{$t=5$}
\end{minipage}
\qquad
\begin{minipage}{4.8cm}
\includegraphics[width=4.8cm]{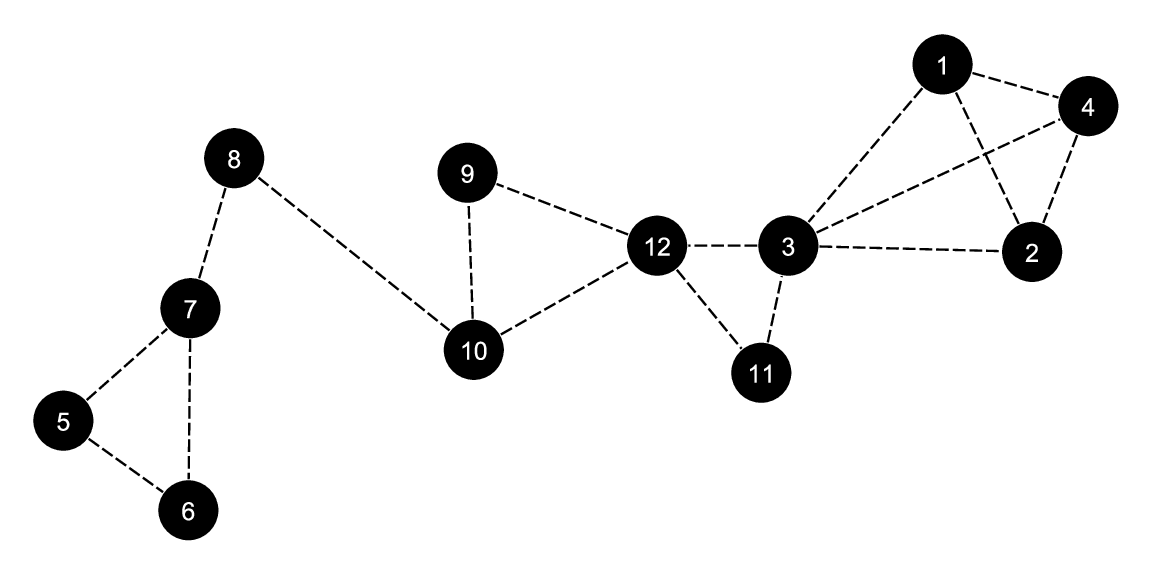}
\caption*{$t=6$}
\end{minipage}
\caption{An illustration of the model. The color of the nodes represents how likely a node will leave in the future, where white nodes are very unlikely to leave and the level of grayness correlates with the probability to leave. Whenever a node leaves the network it is marked as black, all its edges are removed, and all of its neighbors get affected by its leave by increasing their leave probability. The dotted edges are the removed edges.}
\label{fig:toy1}
\end{figure}
\vspace*{-0.2cm}
\subsection{Probability Gain}
\label{sec:probabilitygain}
At any point of time time $t$ where $t \textgreater 0 $, the node leave probability is changed from $\pi_w^{t-1}$ to $\pi_w^{t}$ depending on the leave of $w$'s neighbors. Assume that a node $w$ did not leave the network at time $t$, then we have two sets:
\begin{itemize}
\item $\overline{\Gamma}^{t-1}_w$: the set of $w$'s neighbors who left the network at $t-1$.
\item  $\underline{\Gamma}^{t-1}_w$: the set of $w$'s neighbors who did not leave the network at $t-1$.
\end{itemize}

Thus, at any point of time $t>0$ we have: $|\overline{\Gamma}^{t}_w| = |\Gamma_w^{t-1}| - |\underline{\Gamma}^{t}_w|$ for a decaying network. 
We assume that in the online social settings that when many neighbors (friends) of a node $w$ leave the social network at time $t-1$, then the probability that the node $w$ will leave at $t+1$ will increase due to the probability gain the node $w$ gained at $t$. This difference of the probability between the time $t-1$ and the time $t$ is called \textit{Probability Gain}, $\Delta \pi_w^{t}$.
Thus, a node $w$ will leave at time $t+1$ with probability $\pi_w^{t+1}$ such that:

\begin{equation}
\label{deltapi_original}
\pi_w^{t+1} = min\{1,\ \pi_w^{t-1} + \Delta \pi_w^{t}\}
\end{equation}
We model the probability gain as directly proportional to the number of neighbors who left the network as follows: 
\begin{equation}
\label{eq_directPropotion}
\Delta\pi_w^{t} \propto  |\overline{\Gamma}^{t-1}_w|
\end{equation}
for $t \textgreater 0$.
\subsubsection{Probability gain due to one node leave:}
In order to capture the probability gain meaningfully in the model, we first provide the probability gain due to the leave of one neighbor $v$ of the a node $w$ at time point $t-1$, and then generalize it to the all of left neighbors $\overline{\Gamma}^{t-1}_w$.
In social networks, the interaction between members is not a constant value, different members interact with each other with different intensity. We capture this property in our model by introducing $\delta_{v,w}^{t-1} \in (0,1]$, \textit{Tie Strength} between the members $v,w$ at time point $t-1$.
Now, the probability gain that a node $w$ will get at $t$ due to the leave of its neighbor node $v$ at $t-1$ is:
\begin{equation}
\label{eq:gainDueToOneNodeLeave}
\Delta\pi_w^t(v)= 1-(1-\pi_v^{t-1})(1-\delta_{v,w}^{t-1})
\end{equation}
where the edge $e= (v,w) \in E(G)_{t-2}$ and $e= (v,w) \notin E(G)_{t-1}$ as $ v \in \overline{\Gamma}_w^{t-1}$ and $w \in V(G)_{t-1}$.

Equation~\ref{eq:gainDueToOneNodeLeave} can be generalized to calculate the probability gain produced by the leave of a node $v$ to all of its neighbors, see Figure~\ref{fig:toy2} for an illustration, that did not leave at the same time $l(v)$. Thus, we have:
\begin{figure}
\centering
\parbox{4.8cm}{
\includegraphics[width=4.8cm]{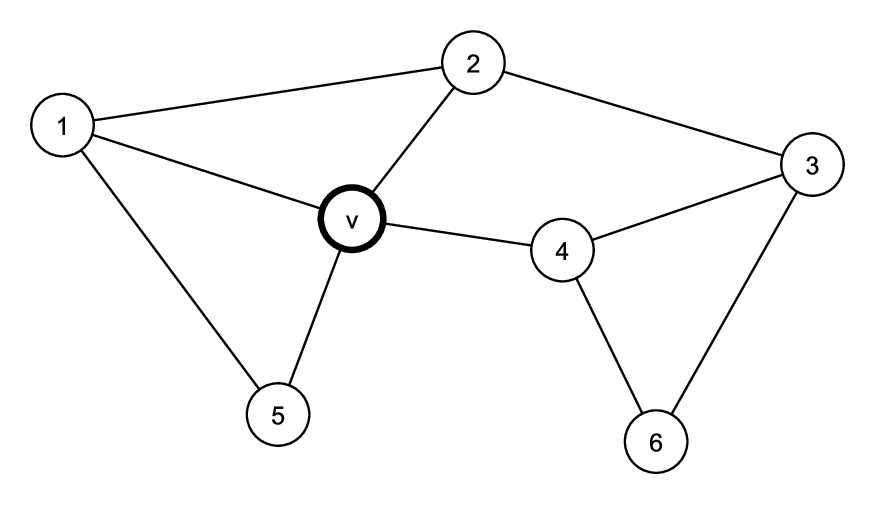}
\caption*{$t-2$}}
\qquad
\begin{minipage}{4.8cm}
\includegraphics[width=4.5cm]{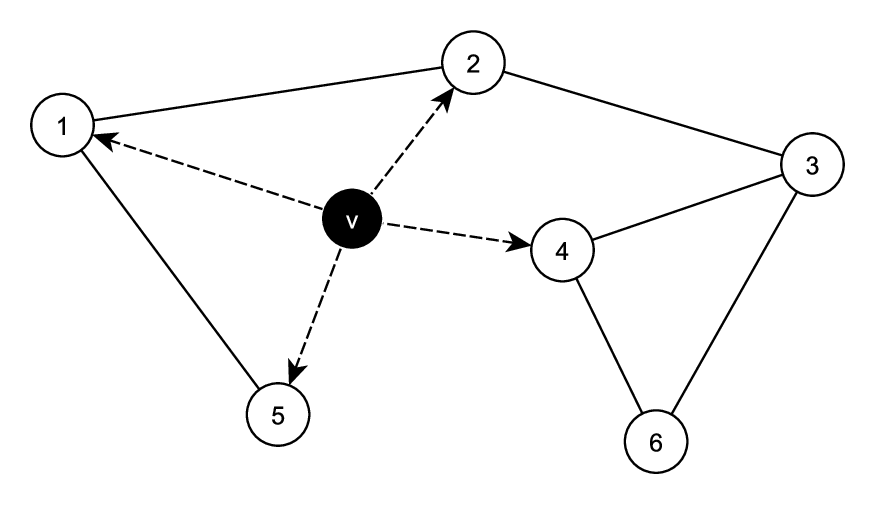}
\caption*{$t-1$}
\end{minipage}
\qquad
\begin{minipage}{4.8cm}
\includegraphics[width=4.8cm]{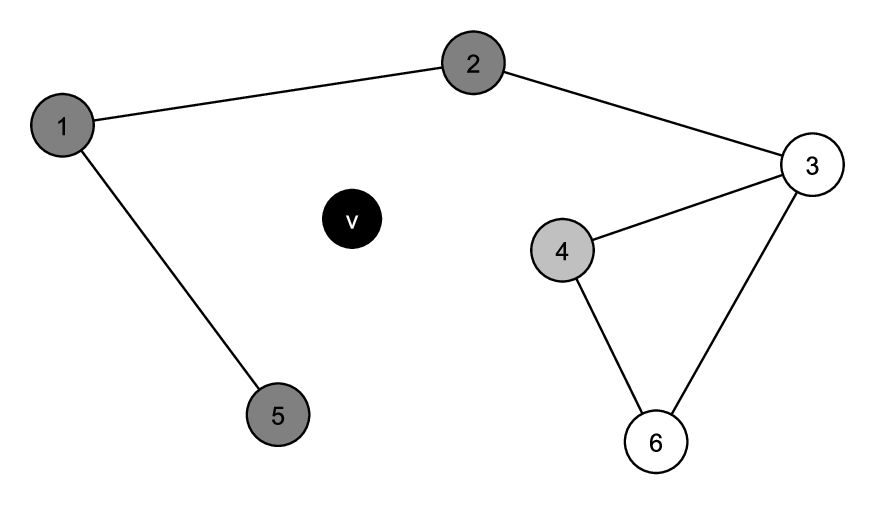}
\caption*{$t$}
\end{minipage}
\caption{This figure shows how a node $v$ affects all of its neighbors when it leaves. At $t-2$, the left network, the node $v$ has a leave probability $\pi_v^{t-2}$ which was gained by $v$'s initial leave probability $\pi_v^0$ and possible probability gain due to leave of its neighbors in earlier time, i.e., $\pi_v^{t-2}=\pi_v^0+\sum_{t=1}^{t=t-3}{\Delta\pi_{v}^{t}}$. At time $t-1$, the middle network, the node $v$ leaves the network affecting its neighbors by increasing the leave probability of nodes $1,2,4,5$. Here we assume that the tie strength between $v$ and the nodes $1,2,5$ is greater than the tie strength between $v$ and $4$. That is why the nodes $1,2,5$ gain more leave probability than node $4$, which is represented in the figure as nodes $1,2,5$ have colors closer to red than node's $4$ color.}
\label{fig:toy2}
\end{figure}

\begin{equation}
\label{eq:nodeLeaveAllGain}
\Delta\pi^t(v) = \sum_{w \in \underline{\Gamma}_v^{t-1}} 1-(1-\pi_v^{t-1})(1-\delta_{v,w}^{t-1})
\end{equation}

\subsubsection{Probability gain due to multiple nodes leave:}
Now, we will formalize what happens when many neighbors of a node $w$ leave the network. We start with a node $w$ that did not leave the network at $t-1$, but some of its neighbors did. So, the leave probability of node $w$ will increase due to the leave of its neighbors, see Figrue~\ref{fig:toy3}, according to the following:

\begin{figure}
\centering
\parbox{4.8cm}{
\includegraphics[width=4.8cm]{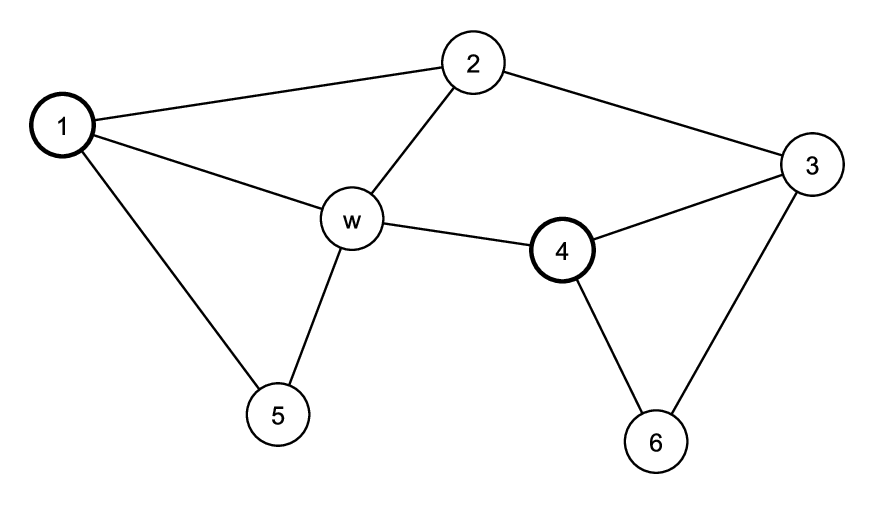}
\caption*{$t-2$}}
\qquad
\begin{minipage}{4.8cm}
\includegraphics[width=4.8cm]{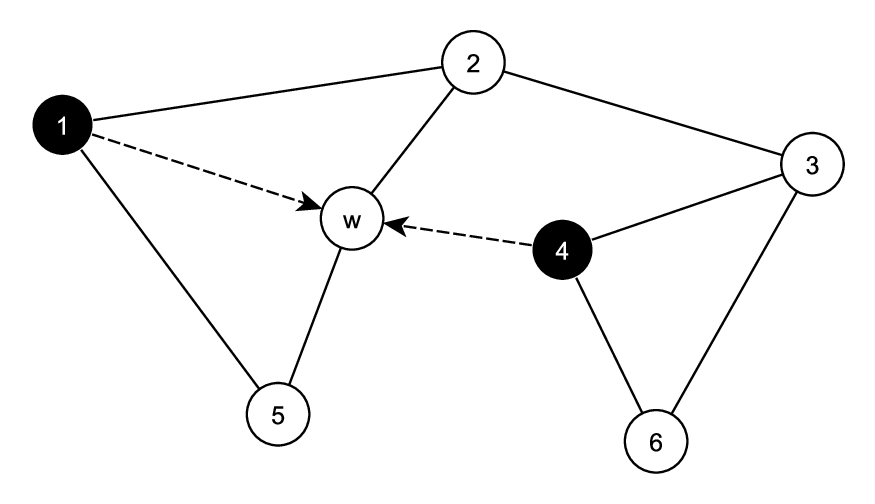}
\caption*{$t-1$}
\end{minipage}
\qquad
\begin{minipage}{4.5cm}
\includegraphics[width=4.8cm]{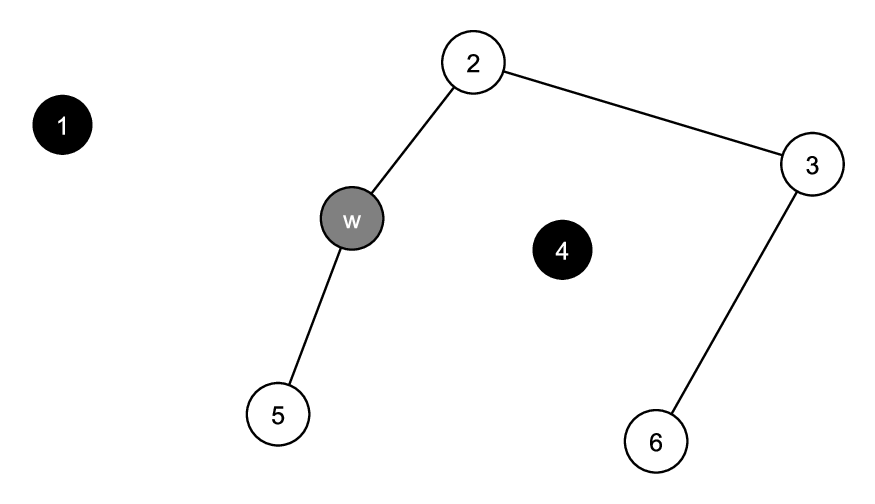}
\caption*{$t$}
\end{minipage}
\caption{This figure shows how a node $w$ is affected by the leave of its neighbors. At $t-2$, the left network, the nodes $1,4$ have leave probabilities $\pi_1^{t-2}$ and $\pi_4^{t-2}$, respectively, which were gained by the node's initial leave probabilities $\pi_1^0$ and $\pi_4^0$ and possible probability gain due to leave of their  neighbors in earlier time. At time $t-1$, the middle network, the nodes $1,4$ leaves the network affecting their neighbors, here we are interested in the node $w$. The leave of nodes $1,4$ left node $w$ with an increased leave probability at $t$. Note that nodes $2,3,5,6$ are affected also by the leave of $1,4$, but for simplicity and for visualization traceability we concentrated on node $w$.}
\label{fig:toy3}
\end{figure}

\begin{equation}
\label{eq:prpbability_gain_with_left_nodes_incorporated_1}
\begin{split}
\Delta \pi_w^{t} &= 1 - [\underbrace{(1-\xi^{t-1}_{w} )}_\text{Assures leave} (\underbrace{\prod_{u \in \overline{\Gamma}^{t-1}_{w}} (1-\pi_u^{t-1}))}_\text{Leave probabilities effect}(\underbrace{\prod_{u \in \overline{\Gamma}^{t-1}_{w}} (1-\delta_{u,w}^{t-1}))}_\text{Tie strength effect}] \\
&= 1 - [(1-\xi^{t-1}_{w} )(\prod_{u \in \overline{\Gamma}^{t-1}_{w}} (1-\pi_u^{t-1})(1-\delta_{u,w}^{t-1}))]
\end{split}
\end{equation}

where $\xi^{t-1}_{w} = \frac{|\overline{\Gamma}^{t-1}_w|}{|\Gamma_{w}^{t-1}|}$ and the quantity $1-\xi^{t-1}_{w}$ assures that when all of the neighbors of the node $w$ leaves, then the node $w$ will (be forced to) leave too as it will be, intuitively, disconnected.

Thus, Equation~\ref{eq:prpbability_gain_with_left_nodes_incorporated_1} becomes:

\begin{equation}
\label{eq:prpbability_gain_with_left_nodes_incorporated_2}
\pi_w^{t} = min\{1,\ \pi_w^{t-1} + 1 - [ (1-\xi^{t-1}_{w} )(\prod_{u \in \overline{\Gamma}^{t-1}_{w}} (1-\pi_u^{t-1})(1-\delta_{u,w}^{t-1}))] \}
\end{equation}

\vspace*{-0.2cm}
\subsection{Node loss}
Let $\vartheta_t$ be the set of nodes that left at time $t$, then the expected number of nodes that will leave the network at $t=1$ is $ |\vartheta_{1}| = \sum\limits_{v \in V}{\pi_w^0} $, and generally we have $ |\vartheta_{t+1}| =|V(G)_t| - \sum\limits_{v \in V}{(1-\pi_w^t)} = \sum\limits_{v \in V}{\pi_w^t} $.
The set of all nodes that left the network between $i$ and $j, \forall i,j > 0$ is: $\vartheta_{i\rightarrow j} = \bigcup\limits_{t\in[i,j]}{\vartheta_t}$.\\
The probability that a node $w$ gets disconnected, i.e., $deg(w)^{t+1}=0$ is:

$P[deg(w)=0]^{t+1} = P[ |\overline{\Gamma}^{t+1}_w| = |\Gamma_w^t|] =\prod\limits_{x\in \Gamma_w^t}{\pi_x^t}$.
\vspace*{-0.2cm}
\subsection{Edge loss}
We also can model the number of edges over time for these networks as follow. Let $\Omega_{t-1}$, \textit{edge loss}, be the set of edges that were removed due to node left at $t-1$, then we have:\\

\begin{equation}
\label{edgesLoss}
|\Omega_{t-1}| =\sum\limits_{w \in \vartheta_{t-1}}{deg(w)} - \sum\limits_{\substack{ u,v \in \vartheta_t\\{ e=(u,v) \in E(G^{t-1})}}}1
\end{equation}

Based on this, the expected number of edges that will be removed one step forward in time is:

\begin{equation}
\label{edgeloss_prediction}
\begin{split}
|\Omega_{t+1}| &= \sum\limits_{w \in V(G^t)}{deg(w) \pi_w^t} - \sum\limits_{e=(u,v)\in E(G^t)}{\pi_u^t  \pi_v^t} \\
&= \sum\limits_{e=(u,v)\in E(G^t)}{\pi_u^t + \pi_v^t} - \sum\limits_{e=(u,v)\in E(G^t)}{\pi_u^t  \pi_v^t}\\
&= \sum\limits_{e=(u,v)\in E(G^t)}{\pi_u^t + \pi_v^t - \pi_u^t  \pi_v^t}
\end{split}
\end{equation}

and generally $n$ steps in future, where $n > 0$, as follows:

\begin{equation}
\label{edgeloss_prediction_n}
|\Omega_{t+n}| = \sum\limits_{j=t+1}^{j=t+n}\sum\limits_{e=(u,v)\in E(G^j)}{\pi_u^j + \pi_v^j - \pi_u^j  \pi_v^j}
\end{equation}

The probability that an edge $e=(u,v) \in G^t$ is removed at $t+1$, i.e., $e=(u,v)\not\in E(G^{t+1})$, is:\\
$P[e=(u,v)\not\in E(G^{t+1})] = P[u \not\in V(G^{t+1})\ or\ v \not\in V(G^{t+1})] = \pi_v^t + \pi_u^t - \pi_v^t  \pi_u^t$

\subsection{Node leave influence} 
The leave influence $LI$ is how  much a node affects its neighbors and causes them to leave. Informally, it is the number of $w$'s neighbors who left before the leave of $w$. Assume that a node $w \in \vartheta_t$ that left the network at time $t$, then the $LI$ is calculated by the following equation:

\begin{equation}
\label{leaveinfluence}
LI(w,n) = \sum\limits_{t}^{t=n}{|\overline{\Gamma}^{t}_w|}
\end{equation}

where $n > t$ is greater than the time $t$ at which node $w$ left. The value of $n$ specifies how many steps in time are considered for finding the $LI$. The leave influence is a useful measure that enables us to predict which members of a network will have a bigger effect after their leave.

Another interesting issue is to find the fraction of your friends who unexpectedly left before you leave. Assume that a node $w$ left the network at time $j$, then the fraction of nodes that left the network before the node $w$ left the network at $j$ is the \textit{Neighbors Leave Resilience}(NLR):\\

\begin{equation}
\label{NLR}
NLR(w) = \frac{\sum\limits_{t=0}^{j}{\overline{|\Gamma}_w^t}|}{|\Gamma_w^{t=0}|}=\frac{\sum\limits_{t=1}^{j}{\sum\limits_{u \in \Gamma_w^{t-1}}{\pi_u^{t-1}}}}{deg(w)^{t=0}}
\end{equation}

\section{Monotonicity and submodularity}
In this section, we prove the monotonicity and submodularity properties of the model equations. We start by defining \textit{submodularity}.
\begin{definition}
\label{def:submodularity_definition}
Let $f: 2^V \rightarrow \mathbb{R}_{\geq0}$, where $\mathbb{R}_{\geq 0} = \left\{ x \in \mathbb{R} \mid x \geq 0 \right\}$, be an arbitrary function that maps the subsets $S$ and $T$ to a non-negative real value, where $S\subseteq T \subset V$. Then, the function $f$ is submodular~\cite{krause2012} if it satisfies the following inequality:\\
$f(S\cup\{v\})-f(S) \geq f(T \cup \{v\})-f(T)$, where $v \in V\setminus T$.
\end{definition}

\begin{lemma}[Order preserving of the probability gain sum]
\label{lem:probabilityorderpreserving_sum}
Let $\pi^t = \{\pi_1,\pi_2,\cdots,\pi_n\}$, where $\pi_i \in \pi^t$ and $\pi_i \in (0,1]$. Then we have:\\
 $\sum\limits_{\pi_i\in \pi^t}{\pi_i} \leq \sum\limits_{\pi_i\in \pi^{t+1}}{\pi_i}$ where $\pi^t \subseteq \pi^{t+1}$, and the sets $\pi^t$ and $\pi^{t+1}$ are defined like above.
\begin{proof}
The proof is trivial for both cases, for $\pi^t = \pi^{t+1}$, and for $\pi^{t+1} = \pi^t \cup H$, where $|H| = k$ and $k>0$.   $\square$
\end{proof}
\end{lemma}

\begin{lemma}[Order preserving of the probability gain product]
\label{lem:probabilityorderpreserving_product}
Let $\pi^t = \{\pi_1,\pi_2,\cdots,\pi_n\}$, where $\pi_i \in \pi^t$ and $\pi_i \in (0,1]$. Then we have:\\
 $\prod\limits_{\pi_i\in \pi^t}{\pi_i} \geq \prod\limits_{\pi_i\in \pi^{t+1}}{\pi_i}$ where $\pi^t \subseteq \pi^{t+1}$, and the sets $\pi^t$ and $\pi^{t+1}$ are defined like above.
\begin{proof}
For $\pi^t = \pi^{t+1}$, the proof is trivial.
Now, Assume that $\pi^{t+1} = \pi^t \cup H$, where $|H| = k$ and $k>0$. For $k=1$ the inequality holds as the set $H$ will contain one element whose value is $(0,1]$.
The proof concludes by induction over $k$.  $\square$
\end{proof}
\end{lemma}

\begin{theorem}
\label{submodularity_nodeLeaveAllGain}
The leave probability gain function, Equation~\ref{eq:nodeLeaveAllGain}, is submodular.
\begin{proof}
Assume that a node $w$ left the network and the set $\Gamma_w^{t-1}$ is the set $w$'s neighbors before leaving the network. Using the definition~\ref{def:submodularity_definition} and Equation~\ref{eq:nodeLeaveAllGain}, we prove the theorem by proving the following inequality: $\sum\limits_{u\in S^*} 1-(1-\pi_u^{t-1})(1-\delta_{u,w}^{t-1}) - \sum\limits_{u \in S } 1-(1-\pi_u^{t-1})(1-\delta_{u,w}^{t-1}) \geq
\sum\limits_{u\in T^*} 1-(1-\pi_u^{t-1})(1-\delta_{u,w}^{t-1}) - \sum\limits_{u \in T } 1-(1-\pi_u^{t-1})(1-\delta_{u,w}^{t-1}) 
$\\
where $S^*=S\cup\{v\}$, $T^*=T\cup\{v\}$ and $S\subseteq T \subset \Gamma_w^{t-1}$.
For $S=T$, the equality holds, and now we need to show that the inequality is correct for the case where $S \subset T$.
Simplifying the previous equation, we get:\\

$|S^*| - \sum\limits_{u\in S^*} (1-\pi_u^{t-1})(1-\delta_{u,w}^{t-1}) - |S| + \sum\limits_{u \in S } (1-\pi_u^{t-1})(1-\delta_{u,w}^{t-1}) \geq
|T^*| - \sum\limits_{u\in T^*} (1-\pi_u^{t-1})(1-\delta_{u,w}^{t-1}) - |T| + \sum\limits_{u \in T } (1-\pi_u^{t-1})(1-\delta_{u,w}^{t-1})$ 
\\simplifying the previous inequality we obtain:\\
$1 + \sum\limits_{u \in S } (1-\pi_u^{t-1})(1-\delta_{u,w}^{t-1}) - \sum\limits_{u\in S \cup \{v\}} (1-\pi_u^{t-1})(1-\delta_{u,w}^{t-1})  \geq
1 + \sum\limits_{u \in T } (1-\pi_u^{t-1})(1-\delta_{u,w}^{t-1}) - \sum\limits_{u\in T \cup \{v\}} (1-\pi_u^{t-1})(1-\delta_{u,w}^{t-1}) $\\

Using the fact that the sets $ S \cup \{v\}$ and $T \cup \{v\}$ are larger than the sets $S$ and $T$, respectively, with only one element, namely $v$, we can further simplify the previous inequality to:

$ \sum\limits_{u\in S} (1-\pi_u^{t-1})(1-\delta\pi_{u,w}^{t-1}) - \sum\limits_{u\in S} (1-\pi_u^{t-1})(1-\delta\pi_{u,w}^{t-1})+ \Delta\pi_v^{t-1}(w)  \geq
 \sum\limits_{u\in T} (1-\pi_u^{t-1})(1-\delta\pi_{u,w}^{t-1}) - \sum\limits_{u\in T} (1-\pi_u^{t-1})(1-\delta\pi_{u,w}^{t-1})+ \Delta\pi_v^{t-1}(w)$ 
\\which yields:\\
$  \Delta\pi_v^{t-1}(w) = \Delta\pi_v^{t-1}(w)$.   $\square$
\end{proof}
\end{theorem}
The interpretation of the theorem is that, the more friends a left node has, the more leave probability gain the left node leaves in the network. This suggests that the leave of the nodes with high degrees in a network makes bigger disruptions than the nodes with smaller degrees.

\begin{theorem}
\label{monoton_prpbability_gain_with_left_nodes_incorporated}
The leave probability gain function, Equation~\ref{eq:prpbability_gain_with_left_nodes_incorporated_1}, is monotone, i.e., for a node $w$ we have $\pi_w^t \leq \pi_w^{t+1}$ if the node $w$ did not leave the network at $t+1$.
\begin{proof}
The proof of this theorem follows the proof in Lemma~\ref{lem:probabilityorderpreserving_product}.  $\square$
\end{proof}
\end{theorem}

\begin{theorem}
\label{submodularity_prpbability_gain_with_left_nodes_incorporated}
The leave probability gain function, Equation~\ref{eq:prpbability_gain_with_left_nodes_incorporated_1}, is submodular.
\begin{proof}
Based on the definition in~\ref{def:submodularity_definition}, we want to show that the leave probability function is submodular.
Assume that a node $w \in V(G_t)$ and $v \in \underline{\Gamma}_w^{t-1}$ then we have two cases:
\begin{itemize}
\item $S =T$: For this case we have: $S=T=\bigcup\limits_{j=0}^{t-1}\overline\Gamma_w^{j}$.
\item $S \subset T$: For this case we have: $S=\bigcup\limits_{j=0}^{t-k}\overline\Gamma_w^{j}$  and $T=\bigcup\limits_{j=0}^{t-1}\overline\Gamma_w^{j}$ where $\bigcup\limits_{j=t-k}^{t-1}\overline\Gamma_w^{j} \neq \emptyset$,  $\forall k>1$. If $\bigcup\limits_{j=t-k}^{t-1}\overline\Gamma_w^{j} \neq \emptyset$, then that means there is no node left the network between $[t-k,t-1]$ and hence $S=T$, which yields the first case.
\end{itemize}
Based on this, we need to show that the function $f$, the probability gain function defined in Equation~\ref{eq:prpbability_gain_with_left_nodes_incorporated_1}, is submodular. For the first case, $S =T$, the proof is trivial. In the rest, we proof the second case.
The probability leave gain due to the leave of one node is defined in Equation~\ref{eq:prpbability_gain_with_left_nodes_incorporated_1} can be rewritten as the following in order to prove the submodularity\footnote{We just restricted the probability gain to a set of nodes that is a subset of the left neighbors of $w$.}
\begin{equation}
\label{submodularity_initial_equation}
\Delta\pi_w^{t-1}(S^*)+\Delta\pi_w^{{t-1}}(T)-\Delta\pi_w^{{t-1}}(S)-\Delta\pi_w^{{t-1}}(T^*) > 0 
\end{equation}

where $S^*=S\cup\{v\}$, $T^*=T\cup\{v\}$, and $\Delta\pi_w^{{t-1}}{(H)}$ is the leave probability gain due to the set of nodes $H=\overline{\Gamma}_w^{t-1}$. Not that Equation~\ref{submodularity_initial_equation} can not be $zero$ as $S \neq T$.\\ 
Suppose that an arbitrary function $h(f(\cdot))$ returns the sign of the value of the function $f(x)$. Then, for the gain function in Equation~\ref{eq:prpbability_gain_with_left_nodes_incorporated_2} we have:\\

\begin{equation}
\label{simplification}
h(1 - [(1-\xi^{t-1}_{w} )(\prod_{u \in \overline{\Gamma}^{t-1}_{w}} (1-\pi_u^{t-1})(1-\delta_{u,w}^{t-1}))) = h(1 - (\prod_{u \in \overline{\Gamma}^{t-1}_{w}} (1-\pi_u^{t-1})(1-\delta_{u,w}^{t-1})))
\end{equation}
That is because, in Equation~\ref{eq:prpbability_gain_with_left_nodes_incorporated_2} the quantity $(1-\xi^{t-1}_{w})$ is always greater than $zero$ (because we have $S\subseteq T \subset V$ and $v \in V\setminus T$ ), then it will never change the sign of the entire formula. Thus, we can ignore it safely if we only want to check the sign of the formula. So, we simplify the left hand side of Equation~\ref{submodularity_initial_equation}, benefiting also from the simplification done in Equation~\ref{simplification} to:

\begin{multline*}
h(\Delta\pi_w^{t-1}(S^*)+\Delta\pi_w^{{t-1}}(T)-\Delta\pi_w^{{t-1}}(S)-\Delta\pi_w^{{t-1}}(T^*)) =\\
h(1 - (\prod_{u \in S^*} (1-\pi_u^{t-1})(1-\delta_{u,w}^{t-1})) + 1 - (\prod_{u \in T} (1-\pi_u^{t-1})(1-\delta_{u,w}^{t-1}))\\ - 1 + (\prod_{u \in S} (1-\pi_u^{t-1})(1-\delta_{u,w}^{t-1})) - 1 + (\prod_{u \in T^*} (1-\pi_u^{t-1})(1-\delta_{u,w}^{t-1})))=\\
h((\prod_{u \in S} (1-\pi_u^{t-1})(1-\delta_{u,w}^{t-1})) + (\prod_{u \in T^*} (1-\pi_u^{t-1})(1-\delta_{u,w}^{t-1}))\\- (\prod_{u \in S^*}(1-\pi_u^{t-1})(1-\delta_{u,w}^{t-1})) - (\prod_{u \in T} (1-\pi_u^{t-1})(1-\delta_{u,w}^{t-1})))
\end{multline*}

We can simplify the previous equation further based on the fact that the sets $S^*$ and $T^*$ contain one more element, $v$, than the sets $S$ and $T$, respectively. Thus we get:\\ 
\begin{multline*}
=h(\underbrace{(\prod_{u \in T}(1-\pi_u^{t-1})(1-\delta_{u,w}^{t-1}))(\Delta_w^{t-1}(v)-1)}_\text{Quantity 1} + \underbrace{(\prod_{u \in S}(1-\pi_u^{t-1})(1-\delta_{u,w}^{t-1})) (1-\Delta_w^{t-1}(v))}_\text{Quantity 2})
\end{multline*}
Obviously, Quantity 1 has a negative value and Quantity 2 is a positive value. However, Quantity 2 is greater than Quantity 1 because $|S| < |T|$ and the value of $\Delta_w^{t-1}(v)$ is the same in both quantities, then Quantity 2 minus Quantity 1 is greater than zero, See Lemma~\ref{lem:probabilityorderpreserving_product}.   $\square$
\end{proof}
\end{theorem}

The theorem state that the more of your friends leave, the less effect the others who stay have on you. I.e., The more of your friends leave, the less important the others become. Submodulariy entails an interesting properties: the minimization problem of submodular function can be performed in polynomial time~\cite{iwata2001}, and the maximization problem of the submodular function (which is NP-Hard problem) can be approximated within a factor of $\alpha =(1-1/e)$ using a greedy algorithm~\cite{nemhauser1978}.
In the following we formalize the optimization problem of the model.

\section{Model optimization}
\subsection{Maximization problem}
The maximization problem under the settings of the model is defined as: Select a set of nodes $\mathcal{A}$ of maximum size $k$ such that the number of left nodes at time $t+1$ is maximum. Equation~\ref{eq:maximization} shows this definition formally.

\begin{equation}
\label{eq:maximization}
     \begin{aligned}
      \text{\textit{Maximize}\ } & |\vartheta_{t+1}|  \\
      \text{\textit{Subject to}}\ & |\mathcal{A}| \leq k, \mathcal{A} \subseteq V(G) &\\
     \end{aligned}
\end{equation}

Intuitively, the number of left nodes is directly proportional to the probability gain. Thus, and based on Equation~\ref{eq:nodeLeaveAllGain}, the maximization problem in Equation~\ref{eq:maximization} becomes:

\begin{equation}
\label{maximization_detaild}
     \begin{aligned}
      \text{\textit{Maximize}\ } & \sum_{v \in V(G)_{t}}\Delta\pi^t(v)= \sum_{v \in V(G)_{t}} \sum_{w \in \Gamma^t_v} 1-(1-\pi_v^{t-1})(1-\delta_{v,w}^{t-1})  \\
      \text{\textit{Subject to}}\ & |\mathcal{A}| \leq k, \mathcal{A} \subseteq V(G^t) &\\
     \end{aligned}
\end{equation}
The previous definition is \textit{one-step} optimization, which means we maximize for a one step in the future only.

\subsection{Minimization problem}
Conversely, the minimization problem under the settings of the model is defined as: Select a set of nodes $\mathcal{A}$ of maximum size $k$ such that the number of left nodes at time $t+1$ is minimum. Equation~\ref{eq:minimization_detaild} shows this definition formally.
\begin{equation}
\label{eq:minimization_detaild}
     \begin{aligned}
      \text{\textit{Minimize}\ } &  \sum_{v \in V(G)_{t}} \sum_{w \in \Gamma^t_v} 1-(1-\pi_v^{t-1})(1-\delta_{v,w}^{t-1})  \\
      \text{\textit{Subject to}} & |\mathcal{A}| \leq k, \mathcal{A} \subseteq V(G) &\\
     \end{aligned}
\end{equation}

\section{The potential of the model}
Having described the model and proved some of its properties, especially the submodularity and the optimization implication, there are different applications.
\begin{itemize}
\item \textit{Leave cascade detection}: a single member leave is not as harmful as a cascade of leaves for the networks that seek growth dynamics. The presented model captures the dynamics of cascade leave by tracking the leave probabilities of the nodes and their increase. 
\item\textit{Maximizing the leave effect:} for a network where a dissolving (disruption) process is required, like criminal social networks, the model is able to provide an accepted level of disruption maximization (thanks to the submodularity property of the model) to the network with insights about the influential members and the effect of the leave of each one.
\item \textit{Social network resilience:} the resilience against huge disruptions, like \textit{Churn Activities} particularly in game industry, in social networks is not well studied. We think that the model provides a first step towards engineering a resilient social network via understanding the decay dynamics of a network.
\end{itemize}

\section{Challenges and future work}
We aim at providing a simple model, and the simplicity imposes some challenges. One of these challenges is estimating the initial leave probabilities. It is not clear whether these probabilities exist in reality or not, however, some analysis like that shown in the Figure~\ref{fig:active_days} for the number of active days before being inactive, may give some insights regarding how to model the initial leave probability.
\vspace*{-0.5cm}
\begin{figure}
\centering
\includegraphics[width=8.5cm]{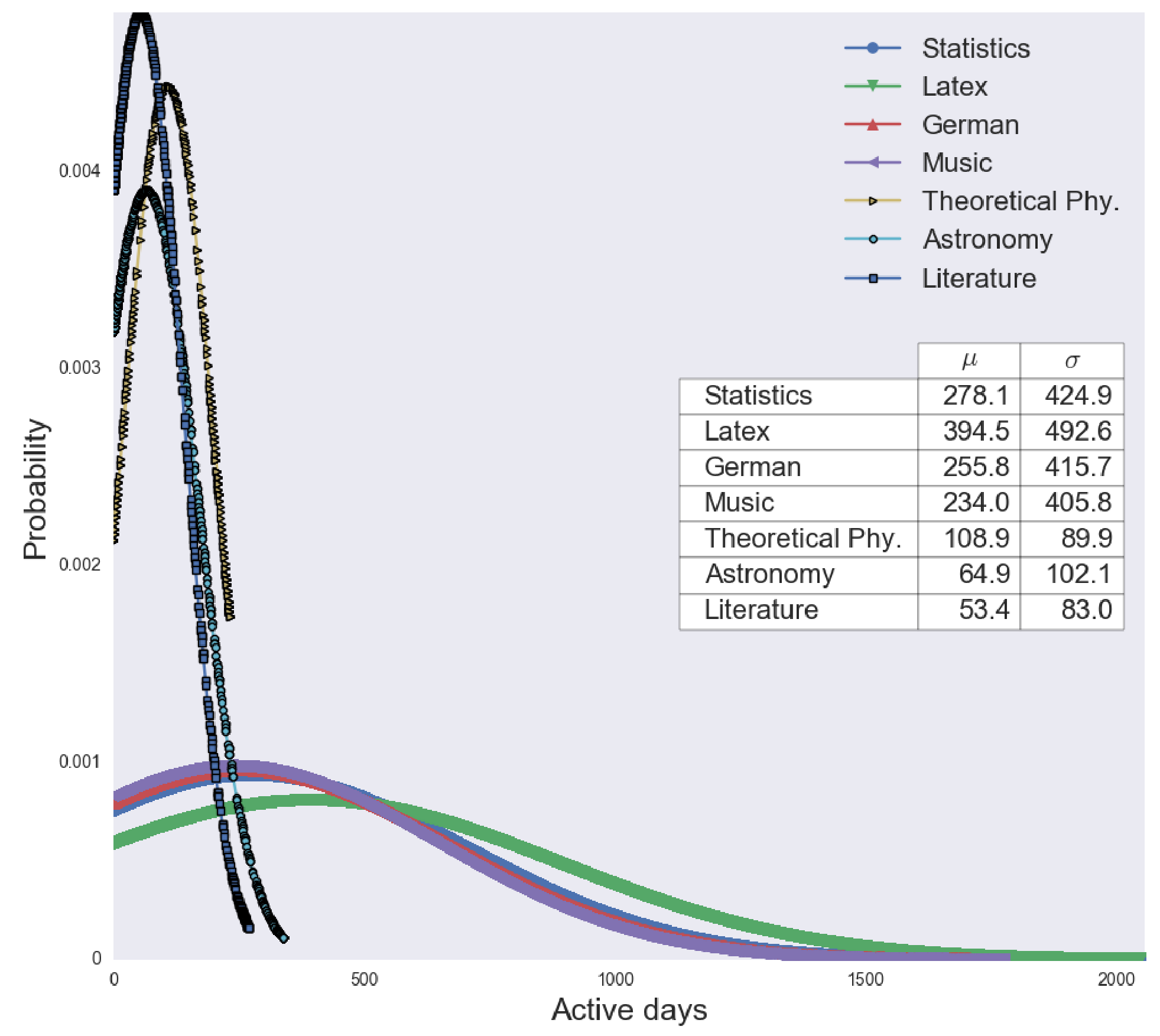}
\caption{(Color Online) The probability distribution of the number of users' active days for different websites. Markers with bold boarders are decayed websites, $\mu$ is the mean, and $\sigma$ is the standard deviation.} 
\label{fig:active_days}
\end{figure} 
The model's intended scope are social networks where edges between members denote common interest and where the time interval is so short that it can be assumed that these do not change within the observed interval. In this case, only the inactivity of nodes (not of edges) needs to be modeled. However, future step is to test and the compare the results of the model with the Stack Exchange data sets as it is the only data set with social decay information we have at the moment.

\section{Conclusion}
In this work, we have presented empirical analysis for the social decay dynamics from the closed Stack Exchange websites. Then, we have presented our model for capturing the decay dynamics in social networks. The model is a probabilistic model that assumes that the leave of social network members is affected by the leave of their neighbors. In this work we have also presented some mathematical proprieties and proved them. We proved that the model main equations are submodular, which entails doing optimization under the setting of the model in feasible way. In the future, we will test the model against the real data sets from the closed Stack Exchange websites and investigate its potentials in understanding and interpreting the mechanics of the online social decay.

\bibliographystyle{ieeetr}
\bibliography{references}
\end{document}